\documentclass{article}
\usepackage{epsf}
\usepackage{amssymb}
\newcommand{\bea}{\begin{eqnarray}}
\newcommand{\eea}{\end{eqnarray}}
\newcommand{\arctanh}{\rm arcth}

\title{
Investigation of ferroelectric order-disorder type compounds with
asymmetric double-well potential
}

\author{
R.R.Levitskii
\thanks{Institute for Condensed Matter Physics, 1 Svientsitskii st., 
79011, Lviv, Ukraine 
\newline \indent \indent
e-mail: werch@icmp.lviv.ua},
T.M.Verkholyak$^*$,
I.V.Kutny
\thanks{Lviv commercial academy,
10 Tugan-Baranovski st., 79005. Lviv, Ukraine},
I.G.Hil'
\thanks{Lviv Ivan Franko National University,
1 Universytetska st., 79000, Lviv, Ukraine}
}
\date{June 18, 2001}
\begin{document}
\begin{titlepage}
\maketitle
{\abstract
{The Mitsui model for the order-disorder type ferroelectrics is 
studied within the mean field approximation. The phase diagram of the 
model is obtained, its dependence on the transverse field is obtained. 
A scheme of setting the values of the model parameters is 
proposed; all the dielectric characteristics of Rochelle salt 
are calculated with the found parameters. Within the Mitsui model 
we study the  physical characteristics of RbHSO$_4$ type crystals; the 
found values of the model parameters provide a good description of both 
dielectric and thermodynamic properties of these compounds.}}
\thispagestyle{empty}
\end{titlepage}

\newpage

\section{Introduction}

The Mitsui model proposed \cite{mitsui} for description of Rochelle salt 
\cite{iona}
 turned out to be suitable for description of the 
 dielectric properties of other ferroelectric crystals (for instance of 
 RbHSO$_4$ \cite{alex}  type) as well. This model is based on the 
 assumption that the ordering structure elements move in double-minima 
 potential wells. Due to the existing in these ferroelectrics internal 
electric field, directed to the opposite sides at the  neighboring sites, 
this potential is asymmetric. In the case of Rochelle salt, spontaneous 
polarization is most likely caused by the hydroxyle group (OH)$_5$ 
\cite{iona}, whereas in NH$_4$HSO$_4$ and RbHSO$_4$ 
spontaneous dipole moments are related to ordering of sulphate 
groups  \cite{alex}.

For the first time, the pseudospin formalism for the Mitsui model was used 
in \cite{zeks}. The values of the effective interaction parameters were 
found, which anebled a description of certain characteristics of Rochelle 
salt in the mean field approximation. It should be noted that the Mitsui 
model can have qualitatively different behavior depending on the values of 
the parameters; this dependence was explored in \cite{vax}. Influence of the 
model parameters on its thermodynamic properties was analysed; the phase 
diagram of the model was obtained; a region of possible values of the model 
parameters suitable for Rochelle salt was found. The dielectric properties 
of Rochelle salt were studied in
\cite{kalenik,levit3,levit4,prep} within the mean field approximation; 
however a description of the static dielectric 
susceptibility and the Curie-Weiss constant was not good. In
\cite{ccc,ccc2} the Mitsui model was treated within the two-particle 
cluster approximation; satisfactory results were obtained. On the other 
hand, in \cite{chain1,chain2,chain3} approximations of the 
interacting chains
for the Mitsui model were proposed.

The Mitsui model was also used for description of NH$_4$HSO$_4$ and 
RbHSO$_4$ crystals \cite{alex,review,blat}. Relaxation dynamics of these 
compounds was studied in
\cite{levit3,grigas,levit,levit2,ftt}.
Unfortunately, within the mean field approximation it is impossible to 
obtain such values of the model parameters which would predict the 
low-temperature first order paraelectric-ferroelectric phase transition in 
NH$_4$HSO$_4$ crystals \cite{review}.
It was shown \cite{blat} that the temperature dependence of polarization 
in these crystals can be described with the temperature dependent model 
parameters. A more rigorous theory was developed in \cite{chain3}, where 
exact taking into account of interaction between neighboring chains allowed 
to reproduce the correct temperature dependence of polarization in 
NH$_4$HSO$_4$.

An important problem is to take into account tunneling of the 
ordering elements through the potential barrier (a transverse field in the 
Hamiltonian of the Mitsui model \cite{zeks}. This problem was studied in
\cite{vax,kalenik,mori,zeks2,levsor}.
In \cite{kalenik} the isotopic effect in deuterated Rochelle salt 
was described by changing tunneling only. In
\cite{mori} the spontaneous polarization and static dielectric 
susceptibility of the Mitsui model 
are calculated for several values of tunneling 
integral; a dynamic susceptibility from the Bloch type equation is  
obtained.

It should be mentioned that in Rochelle salt a strong 
piezoelectric effect is present; hence for a correct description of all 
its characteristics we should take into account an interaction of the 
ordering structure elements with a phonon subsystem. The pseudospin-phonon 
interaction was taken into account in \cite{phonon1,phonon2}.

The aim of the  present paper was to develop a fitting procedure for the 
ferroelectrics described by the Mitsui model and to find out whether their 
physical characteristics could be described within the mean field 
approximation.

\section{Thermodynamics of the Mitsui model in the mean field 
approximation}
\setcounter{figure}{0}

Let us consider a Hamiltonian of  ferroelectrics placed in 
external electric field and described by the Mitsui model 
\cite{zeks,vax,kalenik}:
\bea
H\!\!\!\!\!\!&=&\!\!\!\!\!-\frac{1}{2}\sum_{f}\sum_{ij}J_{ij}S_i^z(f)S_j^z(f)
-\sum_{ij}K_{ij}S_i^z(1)S_j^z(2)
\nonumber\\
&-&\!\!\!\!\!\Delta\sum_i(S_i^z(1){-}S_i^z(2))
{-}\mu E\sum_{f}\sum_i S_i^z(f)
{-}\Omega\sum_{f}\sum_i S_i^x(f),
\eea
where $S_i^{\alpha}(f)=\frac12\sigma_i^{\alpha}(f)$
is the $\alpha$-component of the pseudospin operator,
$J_{ij}$ and $K_{ij}$ are interactions between the spins of the same and of 
different sublattices;
$\mu$ is the effective dipole moment of the ordering structure elements;
$\Delta$ is the magnitude of asymmetry of the double-well potential; the 
transverse field $\Omega$ characterises a finite height of the potential 
barrier.

Neglecting quadratic in fluctuations terms, we obtain the 
corresponding Hamiltonian of the mean field approximation \cite{vax}:
\bea
H=\sum_i\left\{
\frac{J+K}{4}\xi^2+\frac{J-K}{4}\sigma^2
-\sum_f\left[E(f)\frac{\sigma_i^z(f)}{2}-\Omega\frac{\sigma_i^x(f)}{2}
\right]
\right\},
\\
E(1,2)=\frac{J+K}{2}+\mu E\mp (\frac{K-J}{2}-\Delta).
\nonumber
\eea
Here $\xi=\langle S_i^z(1)\rangle+\langle S_i^z(2)\rangle$ is the parameter 
of a ferroelectric ordering, whereas
$\sigma=\langle S_i^z(1)\rangle-\langle S_i^z(2)\rangle$
is the parameter of an antiferroelectric ordering;
$J$, $K$ are the Fourier transforms of the interaction constants at ${\bf 
q}=0$.

In order to calculate the partition function, we need to exclude the 
terms in the Hamiltonian containing the operators $\sigma_i^x(f)$. With the 
help of the rotation transformation for the spin operators 
\bea
&&\tilde\sigma_i^z(f)=\sigma_i^z(f)\cos\phi_f
+\sigma_i^x(f)\sin\phi_f,
\nonumber\\
&&\tilde\sigma_i^x(f)=-\sigma_i^z(f)\sin\phi_f
+\sigma_i^x(f)\cos\phi_f,
\eea
where $\sin \phi_f=\frac{\Omega}{\sqrt{E^2(f)+\Omega^2}}$,
$\cos \phi_f=\frac{E(f)}{\sqrt{E^2(f)+\Omega^2}}$,
we get the transformed Hamiltonian 
\bea
H=\sum_i\left\{
\frac{J+K}{4}\xi^2+\frac{J-K}{4}\sigma^2
-\sum_f\sqrt{E^2(f)+\Omega^2}\frac{\tilde\sigma_i^z(f)}{2}
\right\}.
\eea
Now we can easily obtain the free energy per site
\bea
\frac{F}{N}&=&-\frac{1}{N\beta}\ln {\rm Sp} {\rm e}^{-\beta H}
=\frac{J+K}{4}\xi^2+\frac{J-K}{4}\sigma^2
\nonumber\\
&-&
\frac{1}{\beta}\sum_f\ln 2\cosh
\frac{\beta\sqrt{E^2(f)+\Omega^2}}{2},
\eea
where $\beta=\frac{1}{k_BT}$ is the inverse temperature.
In the subsequent calculations it is convenient to use the free energy 
normalized per $\frac{\tilde J+\tilde K}{4}$:
\bea
f=\frac{4}{\tilde J+\tilde K}\frac{F}{N}
=\xi^2-a\sigma^2-t\left[
\ln \left(2\cosh\frac{K_1}{t}\right)
+\ln \left(2\cosh\frac{K_2}{t}\right)
\right].
\eea
Here we introduce the notations
$K_1=\sqrt{(\xi-a\sigma+\gamma+e)^2+\omega^2}$,
$K_2=\sqrt{(\xi+a\sigma-\gamma+e)^2+\omega^2}$,
$a=\frac{K-J}{K+J}$, $\gamma=\frac{2\Delta}{K+J}$,
$t=\frac{4T}{\tilde K+\tilde J}$,
$e=\frac{2\mu E}{K+J}$, $\omega=\frac{2\Omega}{K+J}$;
the parameters with tildes, like $\tilde J=J/k_B$,
are  in temperature units.

Unknown parameters $\xi$, $\sigma$ are determined from the 
condition of extremum of the free energy as a function of $\xi$, $\sigma$. 
This yields the following system of equations
\bea
\xi=\frac{1}{2}\left[
 \frac{\xi-a\sigma+\gamma+e}{K_1}\tanh\frac{K_1}{t}
+\frac{\xi+a\sigma-\gamma+e}{K_2}\tanh\frac{K_2}{t}
\right],
\nonumber\\
\sigma=\frac{1}{2}\left[
 \frac{\xi-a\sigma+\gamma+e}{K_1}\tanh\frac{K_1}{t}
-\frac{\xi+a\sigma-\gamma+e}{K_2}\tanh\frac{K_2}{t}
\right].
\eea
Here it is required that its solution
$(\xi_0,\sigma_0)$ is a saddle point of the function $f$, that is, 
this function should have a maximum in the parameter of an 
antiferroelectric ordering $\sigma$
($\frac{\partial^2 f}{\partial\sigma^2}<0$) and
a minimum in the parameter of a ferroelectric ordering $\xi$
($\frac{\partial^2 f}{\partial\xi^2}<0$) \cite{bs}.

Using the known thermodynamic relations, we can calculate the 
other thermodynamic functions:
entropy
\bea
\!s\!\!\!\!&=&\!\!\!\!\frac{S}{N}=-\frac{1}{N}\frac{dF}{dT}=-k_B\frac{df}{dt}
\!=
\!\ln\! \left(2\cosh\frac{K_1}{t}\right)
{+}\ln\! \left(2\cosh\frac{K_2}{t}\right)
\nonumber\\
&-&\frac{1}{t}\left(
K_1\tanh\frac{K_1}{t}{+}K_2\tanh\frac{K_2}{t}
\right)\!,
\eea
and internal energy
\bea
u=\frac{4}{\tilde J+\tilde K}\frac{U}{N}=f+ts
=\xi^2-a\sigma^2-\sum_fK_f\tanh\frac{K_f}{t}.
\eea
To calculate the specific heat
\bea
c_v=t\frac{d s}{d t}
=t\left\{
\frac{\partial s}{\partial t}
+\frac{\partial s}{\partial \xi}\frac{\partial \xi}{\partial t}
+\frac{\partial s}{\partial \sigma}\frac{\partial \sigma}{\partial t}
\right\}
\nonumber
\eea
we need to solve the system of equations for
$\frac{\partial \xi}{\partial t}$ and
$\frac{\partial \sigma}{\partial t}$:
\bea
2\frac{\partial \xi}{\partial t}
=\left(\frac{\partial \xi}{\partial t}
-a\frac{\partial \sigma}{\partial t}\right)M_1-N_1
+\left(\frac{\partial \xi}{\partial t}
+a\frac{\partial \sigma}{\partial t}\right)M_2-N_2,
\nonumber\\
2\frac{\partial \sigma}{\partial t}
=\left(\frac{\partial \xi}{\partial t}
-a\frac{\partial \sigma}{\partial t}\right)M_1-N_1
-\left(\frac{\partial \xi}{\partial t}
+a\frac{\partial \sigma}{\partial t}\right)M_2+N_2,
\nonumber
\eea
where $M_i=\frac{1}{K_i^2}\left(\frac{K_i^2-\omega^2}{t\cosh^2\frac{K_i}{t}}
+\frac{\omega^2}{K_i}\tanh\frac{K_i}{t}\right)$,
$N_i=\frac{\sqrt{K_i^2-\omega^2}}{t^2\cosh^2\frac{K_i}{t}}$.
Solution of this system reads
\bea
\frac{\partial \sigma}{\partial t}=
\frac{N_2(M_1-1)-N_1(M_2-1)}{(M_1-1)(1+aM_2)+(M_2-1)(1+M_1)},
\nonumber\\
\frac{\partial \xi}{\partial t}=
\frac{N_2(1+aM_1)+N_1(1+aM_2)}{(M_1-1)(1+aM_2)+(M_2-1)(1+M_1)}.
\nonumber
\eea
From this we obtain the specific heat:
\bea
c_v&=&t\left\{ \frac{1}{t^2}\left(
\frac{K_1^2}{\cosh^2\frac{K_1}{t}}+\frac{K_2^2}{\cosh^2\frac{K_2}{t}}
\right)
\right.
\nonumber\\
&-&\left.\frac{(1-a)(N_1+N_2)^2+2a(M_2N_1^2+M_1N_2^2)}
{(M_1-1)(aM_2+1)+(M_2-1)(aM_1+1)}
\right\}.
\eea
Polarization of a crystal is calculated from the following 
thermodynamic relations:
\bea
P=\frac{\partial F}{\partial E}=\mu N \xi.
\nonumber
\eea
Experimentally, polarization per unit volume is measured:
\bea
p=\frac{P}{V}=\frac{\mu\xi}{v},
\eea
where $v$ is the unit cell volume.
Hence, the static susceptibility is obtained in the following form:
\bea
\chi=\frac{dp}{dE}
=\frac{2\mu^2}{v}\frac{1}{k_B(\tilde K+\tilde J)}
\frac{d \xi}{d e},
\eea
where $\frac{d \xi}{d e}=
\frac{M_2(1+aM_1)+M_1(1+aM_2)}{(1-M_1)(1+aM_2)+(1-M_2)(1+aM_1)}$.

If we neglect tunneling ($\Omega=0$), the expressions for the 
thermodynamic quantities and the system of  equations for $\xi$ and
$\sigma$ simplify. The system of equations for the order parameter now 
reads
\bea
\xi=\frac{1}{2}\left[\tanh\frac{K_1'}{t}+\tanh\frac{K_2'}{t}\right],
\nonumber\\
\sigma=\frac{1}{2}\left[\tanh\frac{K_1'}{t}-\tanh\frac{K_2'}{t}\right],
\eea
where $K_1'=\xi-a\sigma+\gamma+e$, $K_2'=\xi+a\sigma-\gamma+e$.

The expressions for the thermodynamic functions simplify as well.
Entropy:
\bea
s{=}\ln\left(2\cosh\frac{K_1'}{t}\right)
{+}\ln\left(2\cosh\frac{K_2'}{t}\right)
{-}\frac{1}{t}[2(\xi+e)\xi{-}2(a\sigma{-}\gamma)\sigma],
\eea
internal energy:
\bea
u=-(\xi+2e)\xi+(a\sigma-2\gamma)\sigma.
\eea

\section{Phase diagram of the Mitsui model}
\setcounter{equation}{0}
\setcounter{figure}{0}

It could be easily verified that the Hamiltonian and thermodynamics
functions of the Mitsui model can be expressed via the dimensionless
parameters $a$, $\gamma$, $t$, $e$, $\omega$. Depending on the values of
these parameters, the model has different types of temperature behavior.

{\bf а) the case $\Omega=0$ ($\omega=0$)}.

In order to explore the phase diagram in this case we use the scheme
proposed by Vaks \cite{vax}. To find possible second order phase
transitions in the system, we expand the system (2.13) in  $\xi$ up to the
linear terms:
\bea
&&\xi=\left(1-\tanh^2\frac{-a\sigma+\gamma}{t}\right)\frac{\xi}{t},
\nonumber\\
&&\sigma=\tanh\frac{-a\sigma+\gamma}{t}.
\eea
We obtain 
a system of equations for temperature of the second order phase
transition. From the second equation we find that
$t=\frac{-a\sigma+\gamma}{\arctanh\sigma}$ and reduce (3.1)
to a single equation for $\sigma$:
\bea
\frac{-a\sigma+\gamma}{\arctanh\sigma}=1-\sigma^2.
\eea
Let us introduce the notation $\sigma=\tanh\alpha$ and rewrite this equation
as
\bea
f(\alpha)=-\gamma+a\tanh\alpha+\alpha(1-\tanh^2\alpha)=0,
\eea
$\alpha$ varies from 0 to $\infty$; the values of
$f(\alpha)$ in the limiting points are
$f(\alpha=0)=-\gamma$, $f(\alpha\to\infty)=-\gamma+a$. Since
\bea
f'(\alpha)=\frac{\partial f}{\partial\alpha}
=(a+1-2\alpha\tanh\alpha)(1-\tanh^2\alpha)
\eea
is a monotonic function, it turns to zero only at one point $\alpha_0$ and
has an extremum at this point.

Let us consider the case $\gamma>a$  ($f(\alpha\to\infty)<0$). The
$f(\alpha)$ is an increasing function  at small $\alpha$ and a decreasing
one at large $\alpha$. Hence, it has a maximum at certain $\alpha_0$. If
$f(\alpha_0)<0$, the model cannot undergo the second order phase
transitions, whereas at $f(\alpha_0)>0$ there could be two phase
transitions in the system. The line on the phase diagram
$(a,\gamma)$, separating these two characteristic types of behavior (with
two phase transitions and with none)
is determined from the following system of equations
\bea
&&f(\alpha_0)=-\gamma+a\tanh\alpha_0+\alpha_0(1-\tanh^2\alpha_0)=0,
\nonumber\\
&&f'(\alpha_0)=(a+1-2\alpha_0\tanh\alpha_0)(1-\tanh^2\alpha_0)=0.
\eea
An equation for this line can be written in implicit form
\bea
\frac{\gamma+\sqrt{\gamma^2+1-a^2}}{2}
\tanh\frac{\gamma+\sqrt{\gamma^2+1-a^2}}{2}
=\frac{a+1}{2}.
\eea
On increasing $a$, when $a=\gamma$ one of the second order phase
transitions disappears, and the system goes to the region with a
single second order phase transition.

At $T=0$ the system can be in two states: the ordered
($\xi=1$, $\sigma=0$) or anti-ordered ($\xi=0$, $\sigma=1$) one. Which of the
two states is realised depends on the corresponding energies
\bea
U(\xi=1,\sigma=0)=1,~~ U(\xi=0,\sigma=1)=a-2\gamma.
\eea
In the region
\bea
a-2\gamma>1
\eea
the system is in the ordered ground state since its energy is the lowest,
whereas in the region
 \bea
a-2\gamma<1
\nonumber
\eea
the system is in the disordered state.

The phase diagram of the model is depicted in fig.3.1. The line $AB$
\epsfxsize=11cm
\begin{figure}[p]
\begin{center}
\leavevmode
{\epsffile{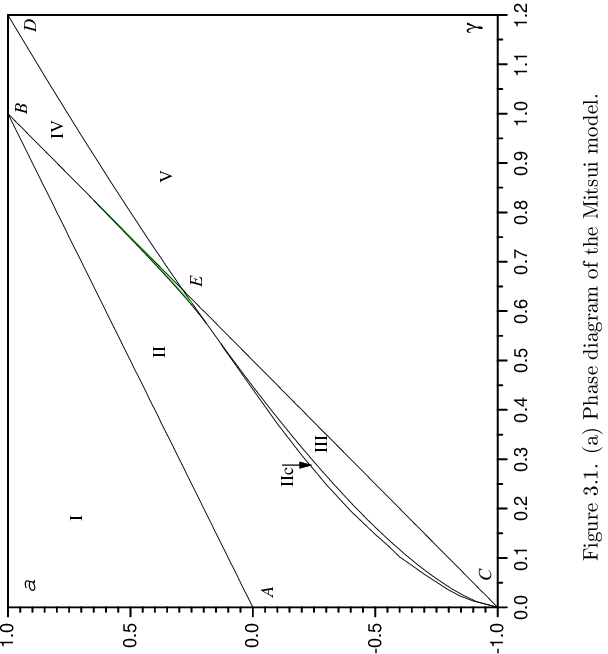}}
\end{center}
\end{figure}
\epsfxsize=11cm
\begin{figure}[p]
\begin{center}
\leavevmode
{\epsffile{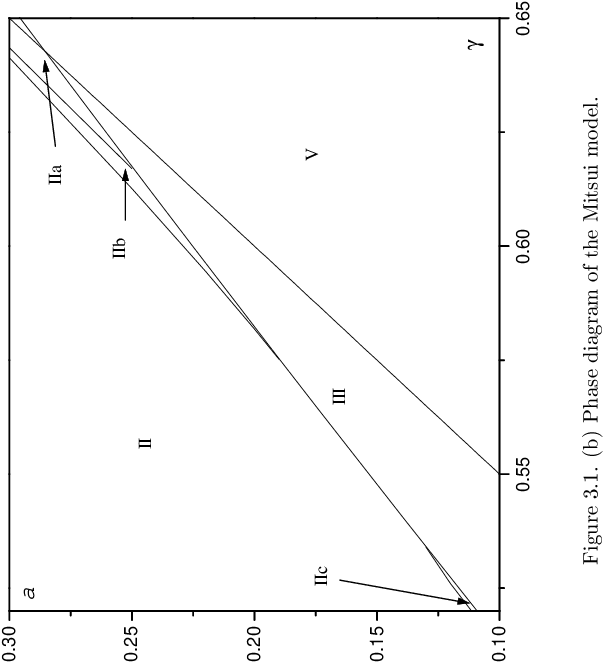}}
\end{center}
\end{figure}
\setcounter{figure}{1}
corresponds to the case $a=\gamma$, $CED$ corresponds to equation
(3.6), $CEB$ corresponds to conditions (3.7) and (3.8). According to
the obtained above results, in the region V of the model parameters values,
the ferroelectric order in the system is absent at any temperature. In the
region IV the ground state of the system in not  ordered ($\xi=0$);
however, here two phase transitions with the order parameter $\xi$ are
possible. This means that $\xi\neq 0$ in a limited interval of
temperatures, as shown in fig.3.2 for $a=0.5$,
$\gamma=0.77$. In the region II the system still can undergo
two second order phase transitions, but its ground state is ordered. In
this region several types of the system behavior are possible; they will be
considered later. In the region I the system undergoes a single first
order phase transition. We fail to obtain simple expressions for temperature
of this transition and restrict our consideration by numerical analysis of
temperature behavior of $\xi$, in order to find new types of the
temperature behavior of the model. The results of this analysis at
different $a$,
$\gamma$ (along the lines $a=const$ at the phase diagram)
are shown in
\epsfysize=6.4cm
\begin{figure}
\begin{center}
\leavevmode
{\epsffile{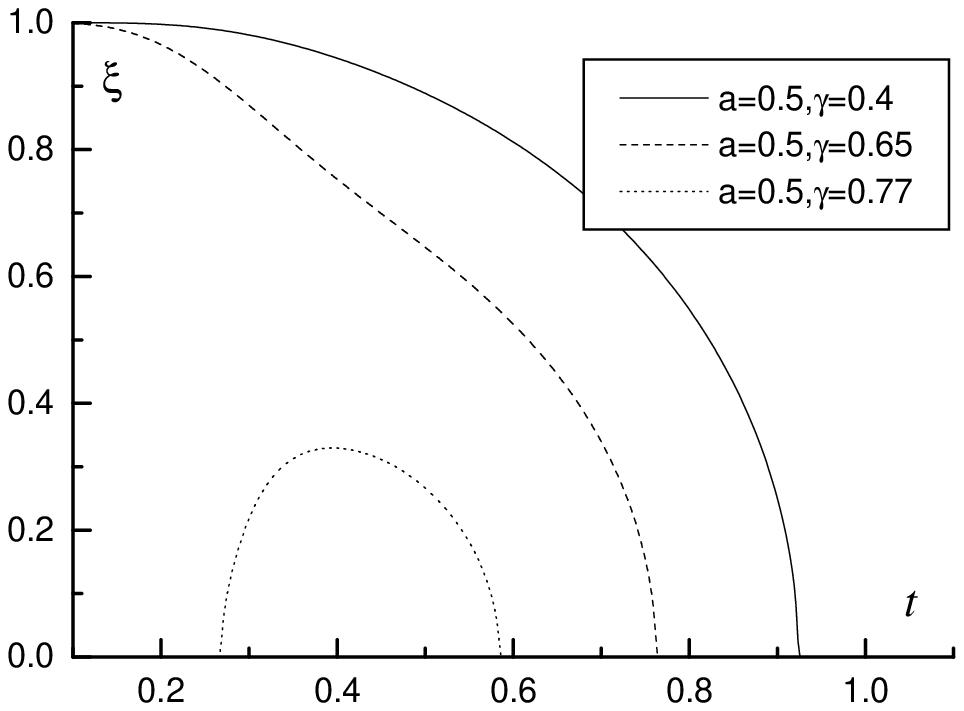}}
\end{center}
\caption{Temperature behavior of $\xi$ at $a=0.5$ and different values of
$\gamma$.}
\end{figure}
\epsfysize=6.4cm
\begin{figure}
\begin{center}
\leavevmode
{\epsffile{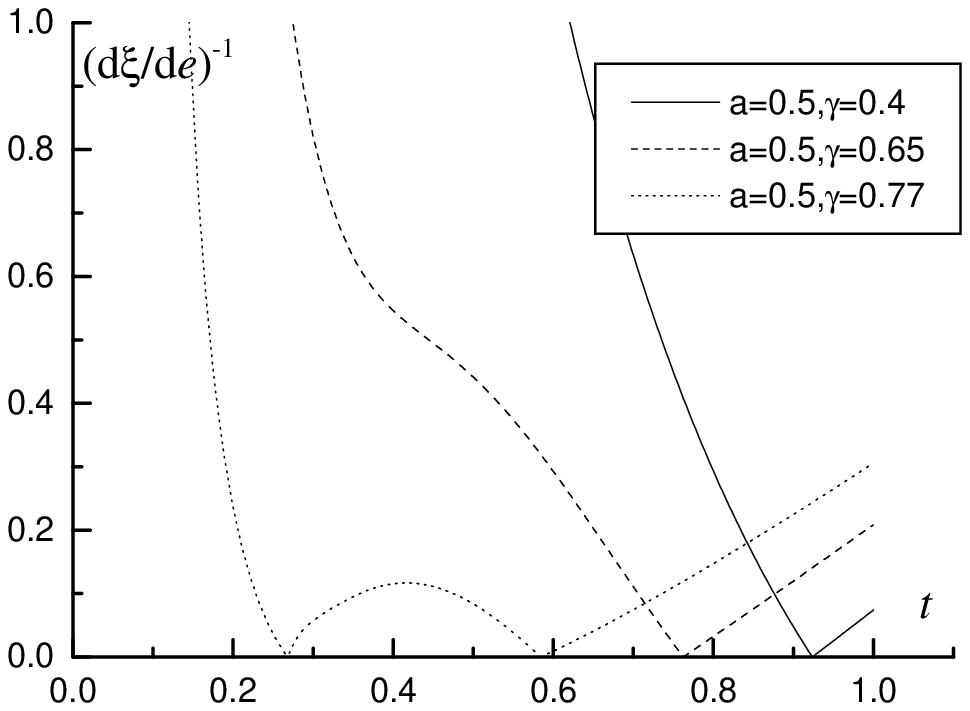}}
\end{center}
\caption{Temperature dependence of $(d\xi/de)^{-1}$
at $a=0.5$ and different values of $\gamma$.}
\end{figure}
\epsfysize=6.5cm
\begin{figure}
\begin{center}
\leavevmode
{\epsffile{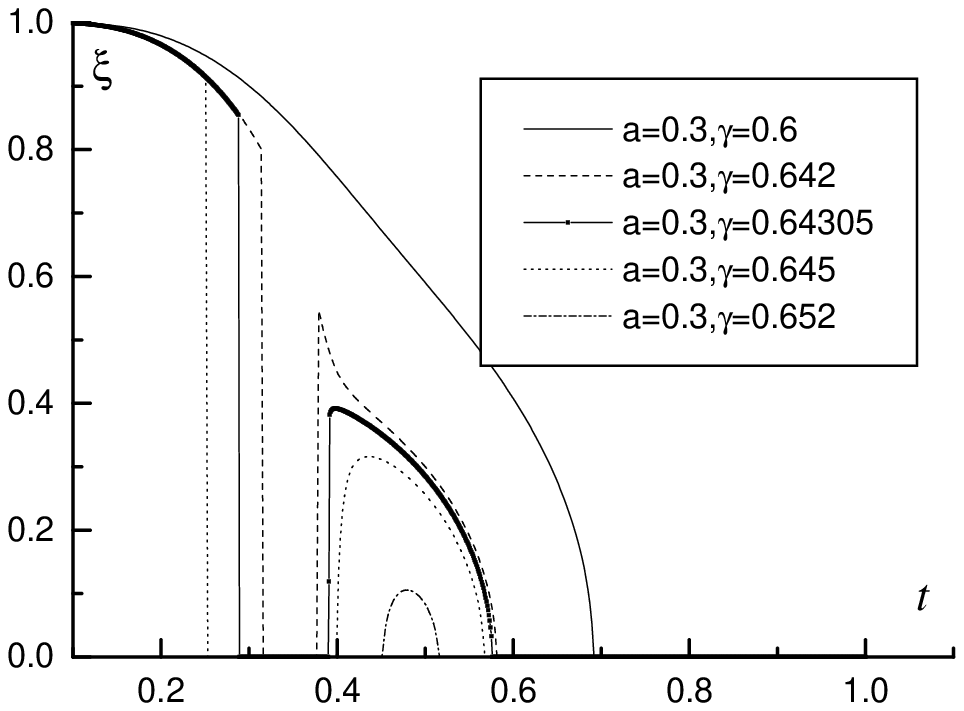}}
\end{center}
\caption{Temperature dependence of $\xi$ at $a=0.3$
and different values of $\gamma$.}
\end{figure}
\epsfysize=6.5cm
\begin{figure}
\begin{center}
\leavevmode
{\epsffile{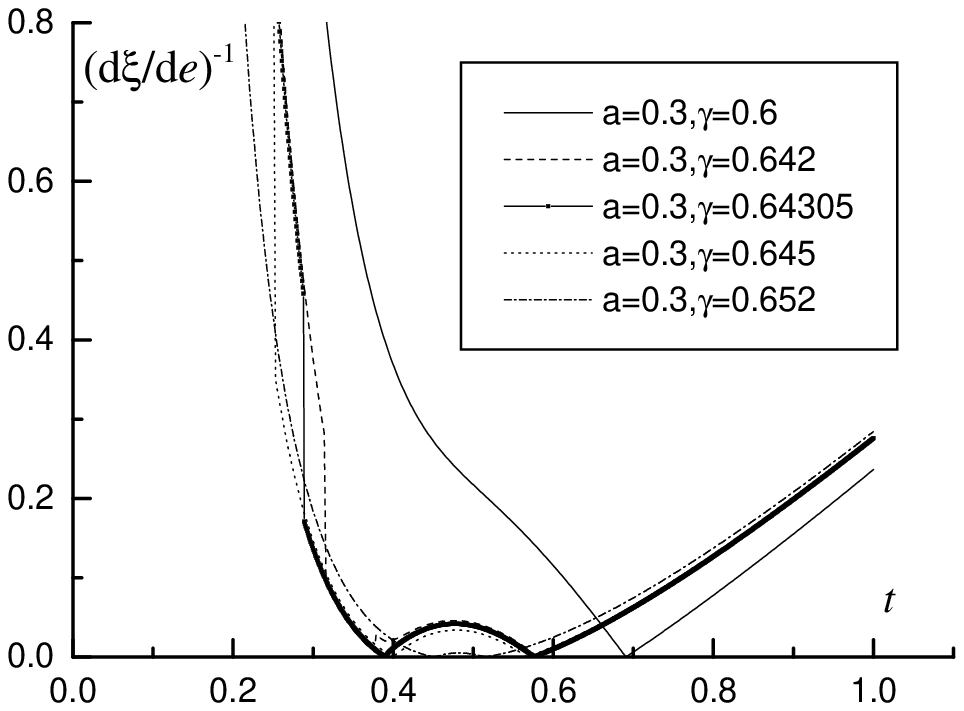}}
\end{center}
\caption{Temperature dependence of $(d\xi/de)^{-1}$
at $a=0.3$ and different values of $\gamma$.}
\end{figure}
\epsfysize=6.5cm
\begin{figure}
\begin{center}
\leavevmode
{\epsffile{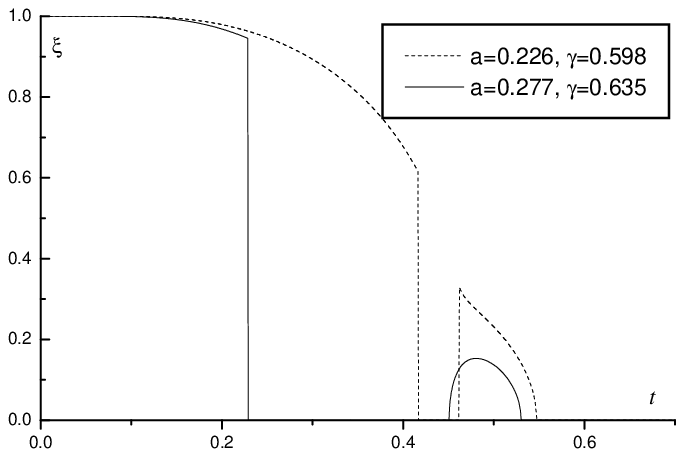}}
\end{center}
\caption{Temperature dependence of $\xi$.}
\end{figure}
\epsfysize=6.5cm
\begin{figure}
\begin{center}
\leavevmode
{\epsffile{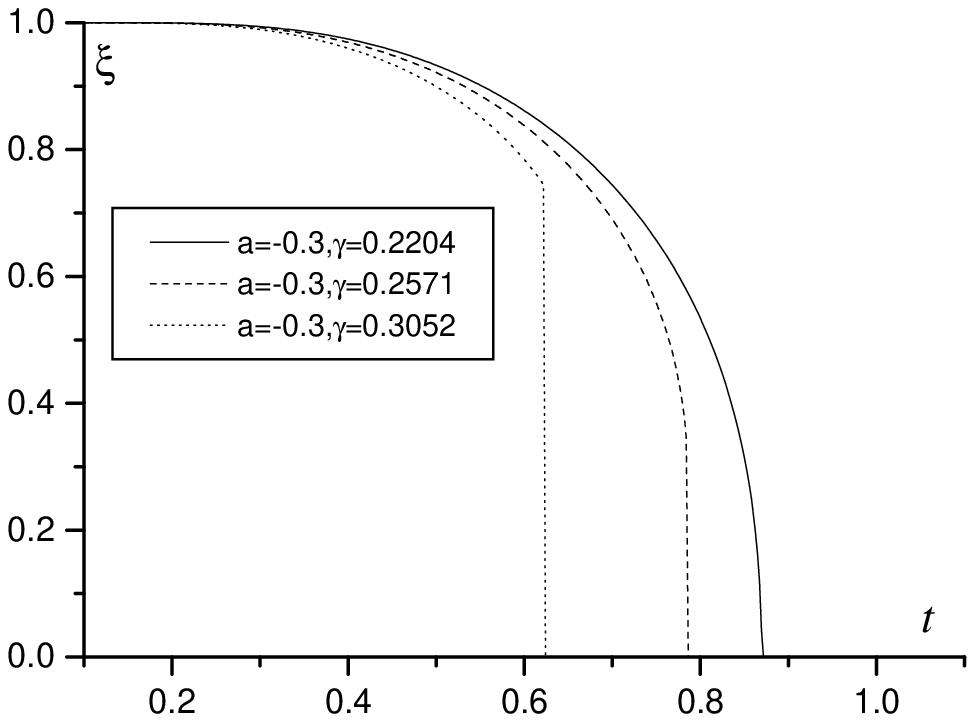}}
\end{center}
\caption{Temperature dependence of $\xi$ at $a=-0.3$
and different values of $\gamma$.}
\end{figure}
\epsfysize=6.5cm
\begin{figure}
\begin{center}
\leavevmode
{\epsffile{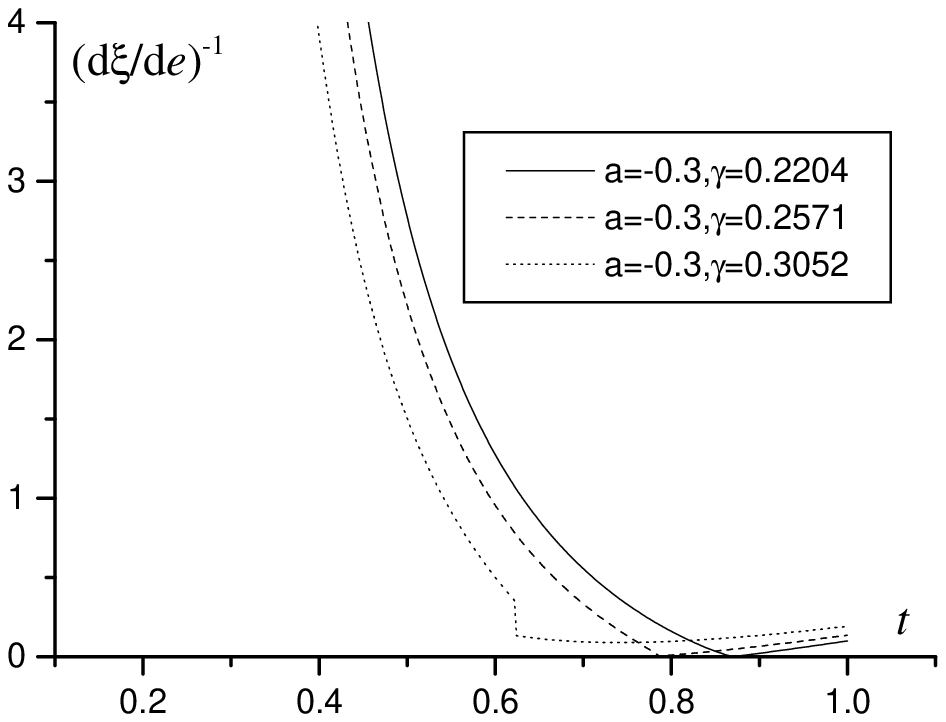}}
\end{center}
\caption{Temperature dependence of $(d\xi/de)^{-1}$
at $a=-0.3$ and different values of $\gamma$.}
\end{figure}
figs.3.2-3.8. We find that in a narrow layer IIa
of the region II there emerges an additional first order phase
transition at low temperature. The temperature dependence of $\xi$ and
other thermodynamic functions of the model in this  region is  shown in
figs.3.4, 3.5 for the curves corresponding to $a=0.3$, $\gamma=0.642$. In
the region IIb the lower second order phase transition
becomes of the first order. Qualitatively, the lines $a=0.3$, $\gamma=0.645$
in figs.3.4, 3.5 illustrates the model behavior in this regions. Deeper in
the region II, the first order phase transition disappears, because the
corresponding transition temperatures become equal ($a=0.3$, $\gamma=0.652$
in figs.3.4, 3.5). On going from the region III to the 
region II, the first
order phase transition persists up to the boundary of the region IIc. The
jump of the order parameter at this boundary turns to zero. At all other
points of the phase diagram, the system undergoes a single second order
phase transition.

{\bf b) the case $\Omega\neq0$.}
As in the case $\Omega=0$, in order to find possible second order phase
transitions at $\Omega\neq 0$
we expand the system (2.7) in $\xi$ up to the linear terms. After some
transformations, it can be presented as
\bea
t&=&\frac{\omega^2t}{[(\gamma-a\sigma)^2+\omega^2]^{3/2}}
\tanh\frac{\sqrt{(\gamma-a\sigma)^2+\omega^2}}{t}
\nonumber\\
&+&\frac{(\gamma-a\sigma)^2}{(\gamma-a\sigma)^2+\omega^2}
\left(
1-\tanh^2\frac{\sqrt{(\gamma-a\sigma)^2+\omega^2}}{t}
\right),
\nonumber\\
\sigma&=&\frac{\gamma-a\sigma}{\sqrt{(\gamma-a\sigma)^2+\omega^2}}
\tanh\frac{\sqrt{(\gamma-a\sigma)^2+\omega^2}}{t}.
\eea
Excluding $t$, we obtain the system (3.9) in the form
$f(\sigma)=0$, where
\bea
f(\sigma)&=&
\frac{\arctanh[\sigma\sqrt{1+\tilde\omega^2}]}{1+\tilde\omega^2}
[1-\sigma^2(1+\tilde\omega^2)]
\nonumber\\
&-&\sqrt{1+\tilde\omega^2}
\left[
(\gamma-a\sigma)-\frac{\tilde\omega^2\sigma}{1+\tilde\omega^2}.
\right].
\eea
Here $\tilde\omega=\omega/(\gamma-a\sigma)$, $\sigma$ changes from 0 up to
$1/\sqrt{1+\tilde\omega^2}$.
Let us find $f(\sigma)$ in the limiting points
\bea
&&f(\sigma=0)=-\sqrt{\gamma^2+\omega^2}<0,
\nonumber\\
&&f(\sigma=\frac{1}{\sqrt{1+\tilde\omega^2}})
=\frac{-\gamma+a\sigma+\sigma(1-\sigma^2)}{\sigma}.
\eea
Hence, the line delimiting the region with a single possible second
order phase transition is a solution of the equation $f(\sigma_{max})=0$.
We obtain the following equation for
$a$, $\gamma$:
\bea
-\gamma+a\sigma+\sigma(1-\sigma^2)=0;
\eea
we also should solve the equation for  $\sigma$:
$\sigma=1/\sqrt{1+\tilde\omega^2}$.
An equation for the line delimiting the regions with two possible second
order phase transitions is obtained from the system of two equations
$\frac{\partial f(\sigma)}{\partial\sigma}=0$, $f(\sigma)=0$. The
expression for the derivative $\frac{\partial f(\sigma)}{\partial\sigma}$
is too cumbersome to be presented here. The equations are solved
numerically; the calculation results are shown in fig.3.9.
\epsfxsize=11cm
\begin{figure}[p]
\begin{center}
\leavevmode
{\epsffile{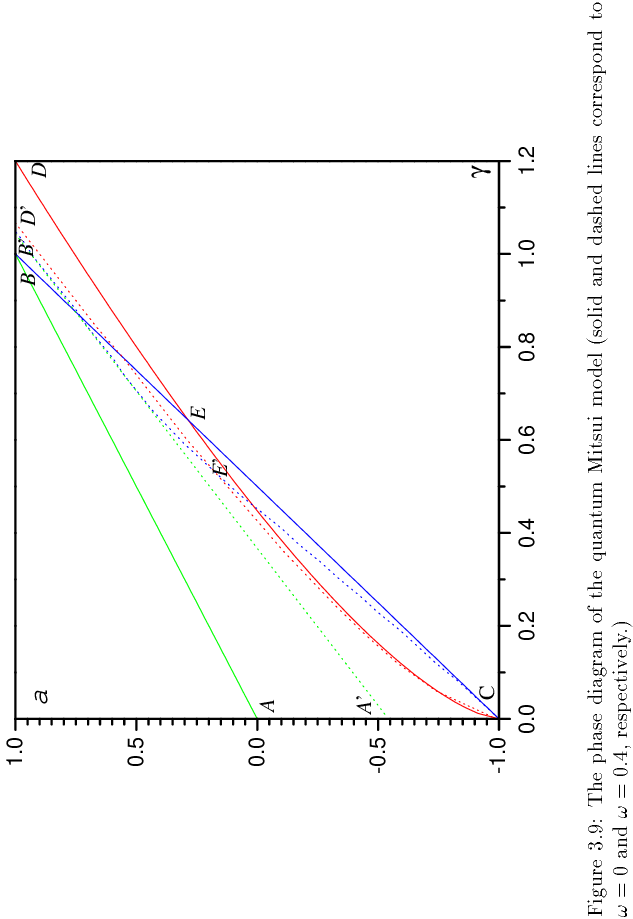}}
\end{center}
\end{figure}
The line separating regions with  ferroelectric ground state and a
disordered ground state is found from the condition
\bea
U(\xi\neq0)=U(\xi=0).
\eea
This equation is also solved numerically. The phase diagram of the
quantum Mitsui model at the value of tunneling $\omega=0.4$ is
shown in fig.3.9 by dashed lines. From this figure one can see that the
boundary of existence of the ordered ground state in raised up and now goes
along the line $CE'D'$. The line of the two phase phase transitions
corresponds to $CE'B'$, whereas the line of a single phase transition
corresponds to $A'B'$.

\section{Thermodynamics of Rochelle salt and deuterated
Rochelle salt.}
\setcounter{equation}{0}
\setcounter{table}{0}
\setcounter{figure}{0}

The ordered phase in the Rochelle salt type crystals exists in a
certain temperature interval: from $T_1=255 K$ to
$T_2=297 K$ for a  regular Rochelle salt and from
$T_1=251 K$ to $T_2=307 K$ for deuterated Rochelle salt.
The transitions to the ordered phase are of the second order in both
crystals. According to the phase diagram (fig.3.1) the possible
parameters for Rochelle salt should be taken from the region IV
\cite{vax,prep}. A difference between the transition temperatures
$\delta_{Rs}=(T_2-T_1)/(T_2+T_1)=0.076$ is small; therefore,
the feasible values of the parameters for Rochelle salt are close to the
line $ED$. In fig.4.1 the line which corresponds to
\epsfxsize=11cm
\begin{figure}[p]
\begin{center}
\leavevmode
{\epsffile{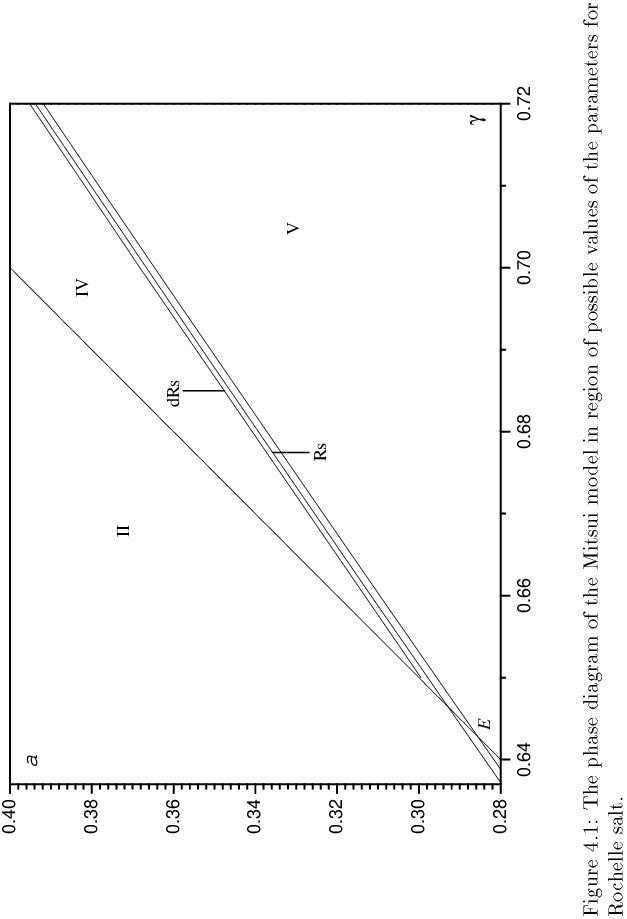}}
\end{center}
\end{figure}
\setcounter{figure}{1}
the sets of the theory parameters for Rochelle salt is depicted; 
above it there lies
a line for deuterated Rochelle salt at which $\delta_{dRs}=0.1$.

The parameters are chosen as follows: for the chosen at the phase
diagram values of $a$, $\gamma$, we find such $K$, $J$, $\Delta$ for which
the theoretical transition temperatures coincide with the  experimental
values; the effective dipole moment $\mu$
is chosen such as the theoretical and experimental  maximal values of
polarization coincide. In figs.4.2-4.5 we plot the examples of
calculated polarization and inverse static dielectric susceptibility for
Rochelle salt and deuterated Rochelle salt at several different sets of the
parameters. One can see, that the best fit for the susceptibility is
obtained at the  small values of $a$ and $\gamma$ from the curves
$Rs$ and $dRs$ (fig.4.1). Hence, the set of the model parameters which
provides the best description of temperature behavior of Rochelle salt
 is: $a=0.29506$, $\gamma=0.648$ ($K=1473.59$~K, $J=802.12$~K,
$\Delta=737.33$~K). The unit cell volume
$v=5.24\times10^{-22}$~cm$^3$ of the model 
is twice smaller than the real one, 
sincethe full unit cell contains four ordering elements
\cite{volume}, whereas the Mitsui model corresponds only to two ordering
elements per unit cell. The effective dipole moment $\mu$
for Rochelle salt is found from the experimental value of
polarization is $\mu=3.04\times10^{-18}$. For the deuterated Rochelle
salt the best set of the fitting parameters is: $a=0.29952$, $\gamma=0.65$
($K=1502.83$~K, $J=8102.07$~K, $\Delta=751.69$~K, $\mu=3.6\times10^{-18}$).

It should be noted that the Rochelle salt was studied in the mean
field approximation by other authors too \cite{vax,kalenik,levit4,prep}.
The parameters values for $Rs$ and $dRs$ obtained in these papers are
given in Table 4.1.

\begin{table}
\caption{The sets of the parameters for regular and deuterated Rochelle
salt obtained in the mean field approximation
(the sets 4 and 5 are found in the present work).}
\begin{tabular}{|c|c|l|l|l|l|l|}
\hline
No &compound           &$K$, K &$J$, K&$\Delta$, K&$a$   &$\gamma$\\
\hline
1. &Rs \cite{kalenik} &1599.84&764.64&815.9      &0.353&0.69    \\
2. &dRs \cite{kalenik}&1563.84&790.85&789.7      &0.328&0.671   \\
3. &dRs \cite{vax}    &1480   &800   &740        &0.298&0.649   \\
4. &Rs                &1473.59&802.12&737.3      &0.295&0.648   \\
5. &dRs               &1502.83&810.07&751.7      &0.299&0.65    \\
\hline
\end{tabular}
\end{table}
\epsfysize=6.2cm
\begin{figure}
\begin{center}
\leavevmode
{\epsffile{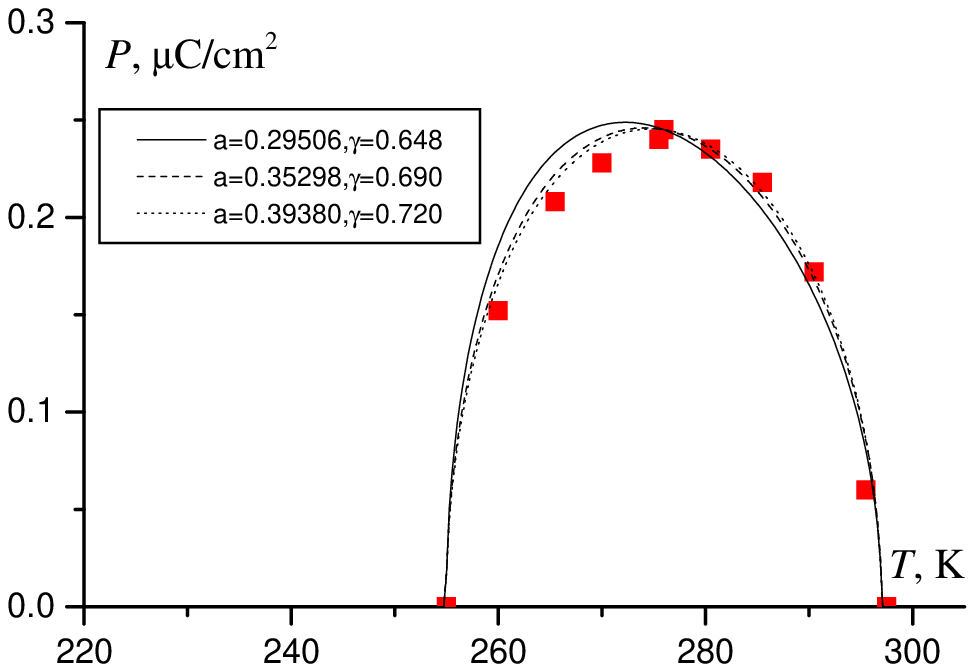}}
\end{center}
\caption{Temperature dependence of polarization of  Rochelle
salt at different sets of the model parameters ($\blacksquare$
are experimental points \protect\cite{iona}).}
\end{figure}
\epsfysize=6.2cm
\begin{figure}
\begin{center}
\leavevmode
{\epsffile{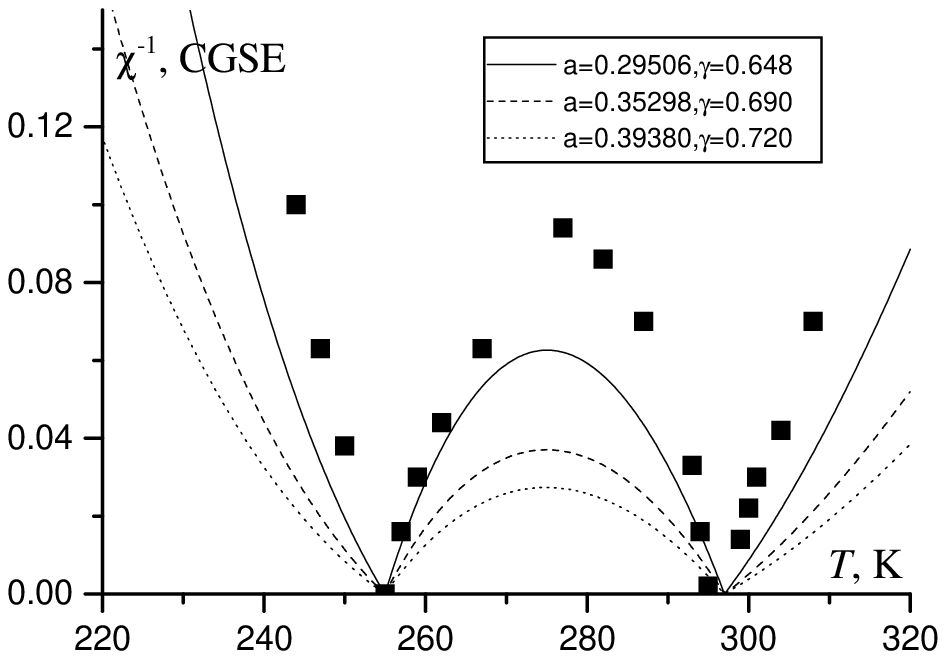}}
\end{center}
\caption{Temperature dependence of inverse static dielectric susceptibility
of  Rochelle salt at different sets of the model parameters ($\blacksquare$
are experimental points \protect\cite{iona}).}
\end{figure}
\epsfysize=6.1cm
\begin{figure}
\begin{center}
\leavevmode
{\epsffile{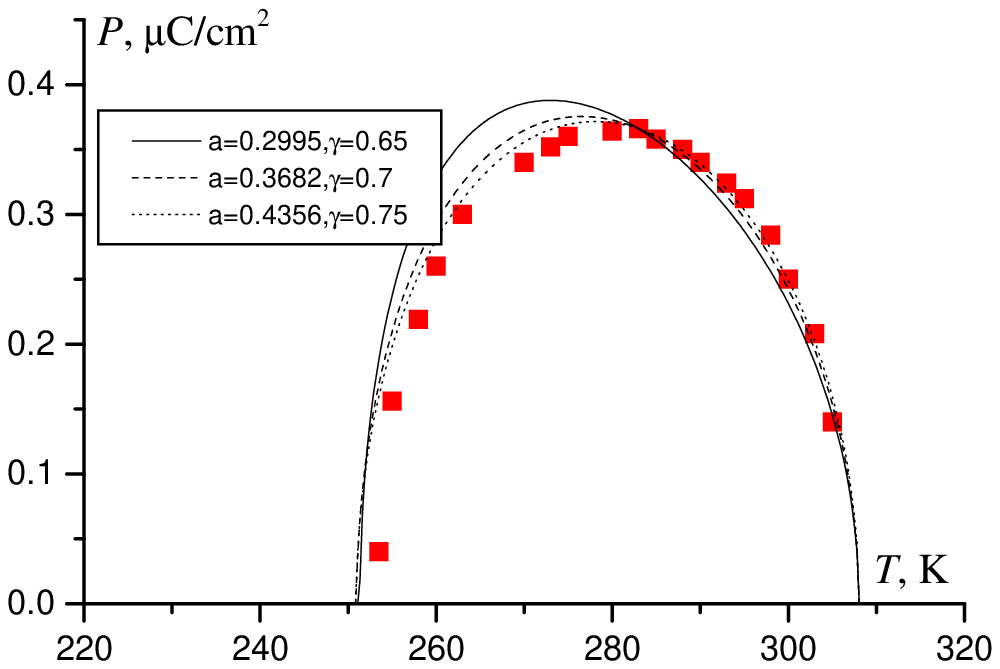}}
\end{center}
\caption{Temperature dependence of polarization of deuterated Rochelle
salt at different sets of the model parameters ($\blacksquare$
are experimental points \protect\cite{iona}).}
\end{figure}
\epsfysize=6.1cm
\begin{figure}
\begin{center}
\leavevmode
{\epsffile{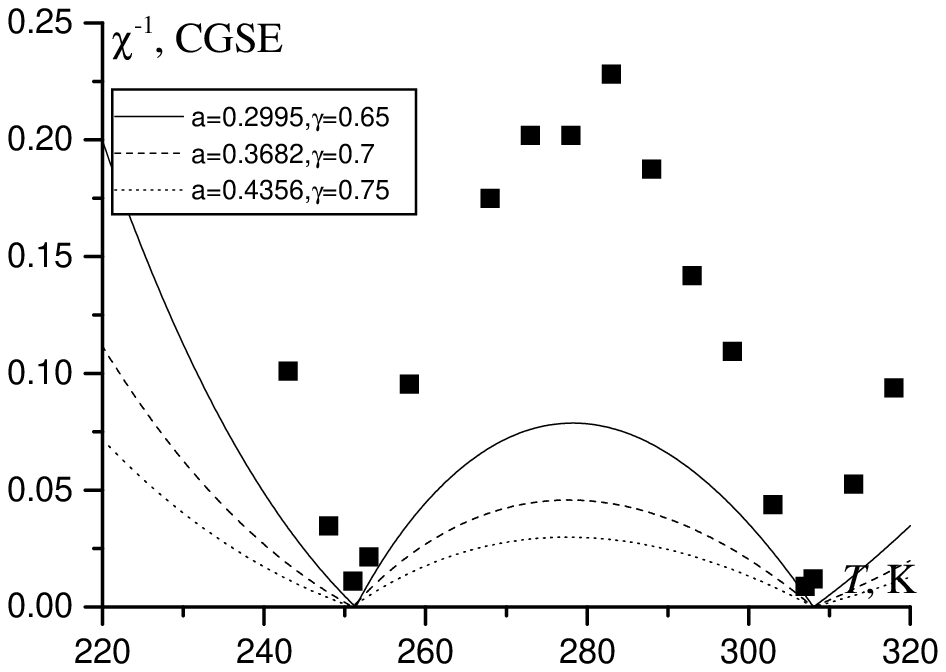}}
\end{center}
\caption{Temperature dependence of inverse static dielectric susceptibility
of  Rochelle salt at different sets of the model parameters ($\blacksquare$
are experimental points \protect\cite{kessenikh}).}
\end{figure}
In order to explore the character of isotopic effect in Rochelle salt, we
studied the influence of transverse field on the thermodynamic properties
of the model. Another fitting procedure for the Rochelle salt was to include
tunneling into the model with the chosen already parameters for deuterated
Rochelle salt, since deuterated mainly decreases the tunneling of the
ordering structure elements. The calculation results are shown in
figs. 4.6, 4.7. They indicate that it is impossible to describe the
present isotopic effect 
by changing the transverse field only.  A thorough
numerical study of transverse field effects on the thermodynamic 
characteristics of the
model has shown that taking into account of tunneling does not improve an
agreement between the theory and experiment. For a fixed $\gamma$, an
increase in $\omega$ only increases the static susceptibility.
\epsfysize=6.0cm
\begin{figure}
\begin{center}
\leavevmode
{\epsffile{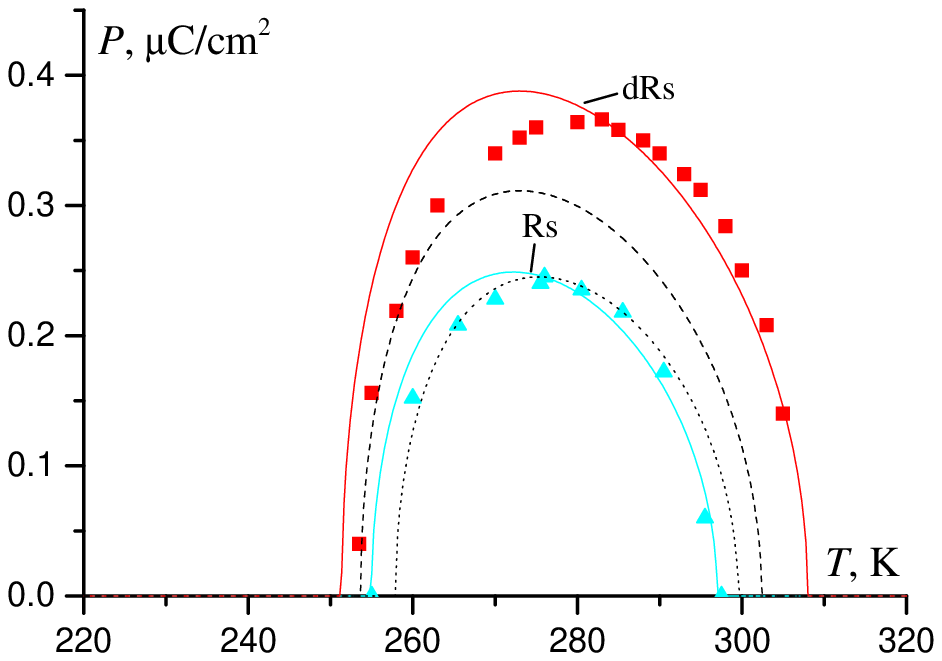}}
\end{center}
\caption{Temperature dependence of polarization of Rochelle salt
and deuterated Rochelle salt for different  sets of the
parameters: lines for Rs and dRs correspond to sets of parameters 4, 5 in 
table 4.1, dashed line is the model with the set of parameters 5 and 
$\Omega=50$.}
\end{figure}
\epsfysize=6.0cm
\begin{figure}
\begin{center}
\leavevmode
{\epsffile{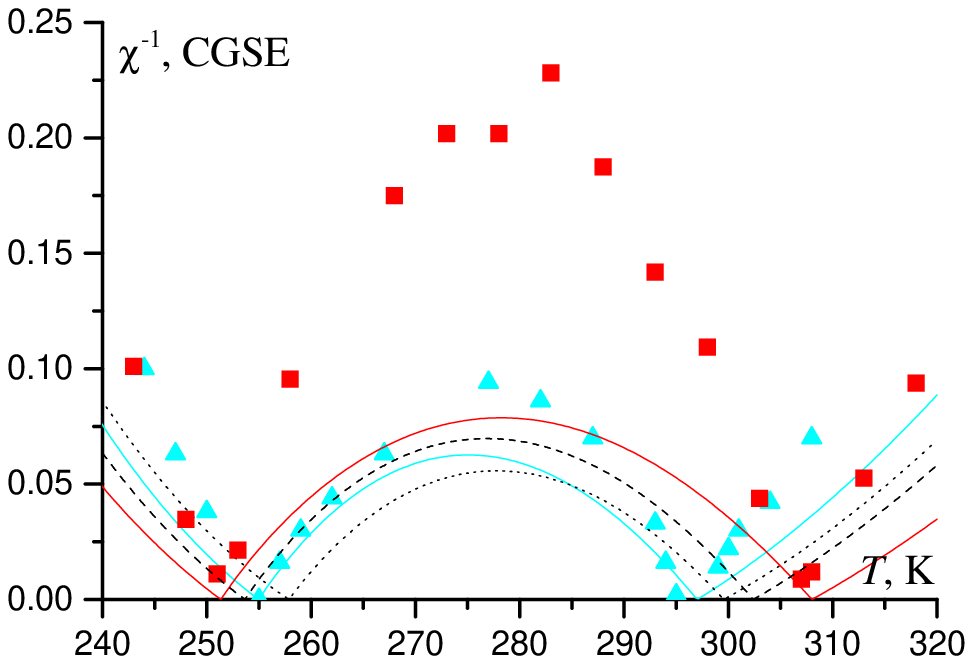}}
\end{center}
\caption{Temperature dependence of inverse static susceptibility
of Rochelle salt and deuterated Rochelle salt for the same sets of the
parameters as in previous figure.}
\end{figure}

The performed calculations of spontaneous polarization and static
dielectric susceptibility for $Rs$ and $dRs$ have shown that it is
impossible to find such sets of the microparameters $K$, $J$, $\Delta$, and
$\mu$, which would provide a simultaneous good description of both
characteristics. Therefore, the statements of other authors that the
Mitsui model is adequate to $Rs$ or $dRs$ and properly describes the
physical characteristics of these crystals are incorrect. Within the model
(2.1) either with tunneling or without it, this is not possible.

\section{Relaxation dynamics of Rs and dRs.}
\setcounter{equation}{0}
\setcounter{figure}{0}

It has been mentioned in several papers \cite{sandy,horioka} that 
as indicated by measurements of dielectric dispersion, tunneling effects 
do not play an important role in Rochelle salt crystals. Therefore, the 
time correlations are believed to be caused by interaction with a phonon 
subsystem. A phenomenologic description of this interaction is performed 
within the Glauber method \cite{glauber}.
A master equation for $\xi$, $\sigma$ in this approach can be presented 
as \cite{prep}:
\bea
-\alpha\frac{d}{dt}\xi=\xi
-\frac{1}{2}\left[
\tanh\frac{K_1'}{t}+\tanh\frac{K_2'}{t},
\right]
\nonumber\\
-\alpha\frac{d}{dt}\sigma=\sigma
-\frac{1}{2}\left[
\tanh\frac{K_1'}{t}-\tanh\frac{K_2'}{t},
\right],
\eea
where $\alpha$ is the parameter of the spin-phonon relaxation.

Considering a weakly non-equilibrium system, we expand $\xi$, $\sigma$ 
near their equilibrium values
${\overline\xi}$, ${\overline\sigma}$:
$\xi={\overline\xi}+\xi(t)$,
$\sigma={\overline\sigma}+\sigma(t)$.
From (5.1) we obtain that
${\overline\xi}$, ${\overline\sigma}$
satisfy  equation (2.13), whereas $\xi(t)$, 
$\sigma(t)$ can be found from the following system of equations
\bea
-\alpha\frac{d}{dt}\xi(t)=M_{11}\xi(t)+M_{12}\sigma(t)-M_1 e(t),
\nonumber\\
-\alpha\frac{d}{dt}\sigma(t)=M_{21}\xi(t)+M_{22}\sigma(t)-M_2 e(t),
\eea
where $M_{11}=1-\frac{1}{t}(1-\xi^2-\sigma^2)$,
$M_{12}=-\frac{2a}{t}\xi\sigma$,
$M_{21}=\frac{2\xi\sigma}{t}$,
$M_{22}=1+\frac{a}{t}(1-\xi^2-\sigma^2)$,
$M_1=\frac{1}{t}(1-\xi^2-\sigma^2)$,
$M_2=-\frac{2\xi\sigma}{t}$.

In a frequency representation, a linear response of $\xi(t)$ to a small 
field $e(t)$ is obtained in the following form
\bea
&&\xi(\omega)=\chi_0(\omega)e(\omega),
\\
&&\chi_0(\omega)
{=}\frac{w_1\tau_1}{1+\omega^2\tau_1^2}
{+}\frac{w_2\tau_2}{1+\omega^2\tau_2^2}
{+}i\omega\left[
\frac{w_1\tau_1^2}{1+\omega^2\tau_1^2}
{+}\frac{w_2\tau_2^2}{1+\omega^2\tau_2^2}
\right],
\\
&&w_1=\frac{M_1(M_11-\frac{\alpha}{\tau_2})+M_2M_{12}}
{\alpha^2(\tau_1^{-1}-\tau_2^{-1})}
\nonumber\\
&&w_2=\frac{M_1(M_11-\frac{\alpha}{\tau_1})+M_2M_{12}}
{\alpha^2(\tau_1^{-1}-\tau_2^{-1})},
\nonumber
\eea
where
$\tau_{1,2}^{-1}=
\frac{M_{11}+M_{22}\pm\sqrt{(M_{11}-M_{22})^2+4M_{12}M_{21}}}{2\alpha}$
are the inverse relaxation times of the model. Relations (5.3), (5.4) for
polarization $P$ and electric field $E$ now read
\bea
&&P(\omega)=\chi(\omega)E(\omega),
\nonumber\\
&&\chi(\omega)=\frac{2\mu^2}{v}\frac{1}{(K+J)}\chi_0(\omega).
\eea
The dielectric permittivity is related to the corresponding 
susceptibility as
\bea
\varepsilon(\omega)=1+4\pi\chi(\omega)=
\varepsilon'(\omega)-{\rm i}\varepsilon''(\omega),
\eea
where
$\varepsilon'(\omega)$ та $\varepsilon''(\omega)$ are the real and 
imaginary parts of the dielectric permittivity.

In figs.5.1-5.6 we depicted the calculated dielectric permittivity
of Rochelle salt. In calculations we use the parameters found above:
$K=1473$, $J=802.12$, $\Delta=737.33$ for Rochelle salt and
$K=1502$, $J=810.07$, $\Delta=751.69$ for deuterated Rochelle salt.
Unfortunately, at the already set values of $\mu$ we are unable to 
find a value of the relaxation parameter $\alpha$ such as to obtain a 
satisfactory description of the experimental points for the dynamic 
permittivity. Therefore, in this case we take $\mu=1.8427\times10^{-18}$
for Rochelle salt and $\mu=2.0\times10^{-18}$ for deuterated Rochelle salt. 
Polarization and static susceptibility calculated with the new 
values of $\mu$ are shown in figs.5.7, 5.8.

As one can see in figs. 5.1-5.6, the mean field approximation 
yields a fair description of the temperature behavior of 
the dynamic dielectric permittivity of Rochelle salt at all 
temperatures except for the vicinity of the phase transitions, where the 
real part of the permittivity tends to $\varepsilon_\infty$. This is due to 
the divergence of the relaxation times of the model at the transition 
temperatures (fig.5.9).
\epsfysize=6.3cm
\begin{figure}
\begin{center}
\leavevmode
{\epsffile{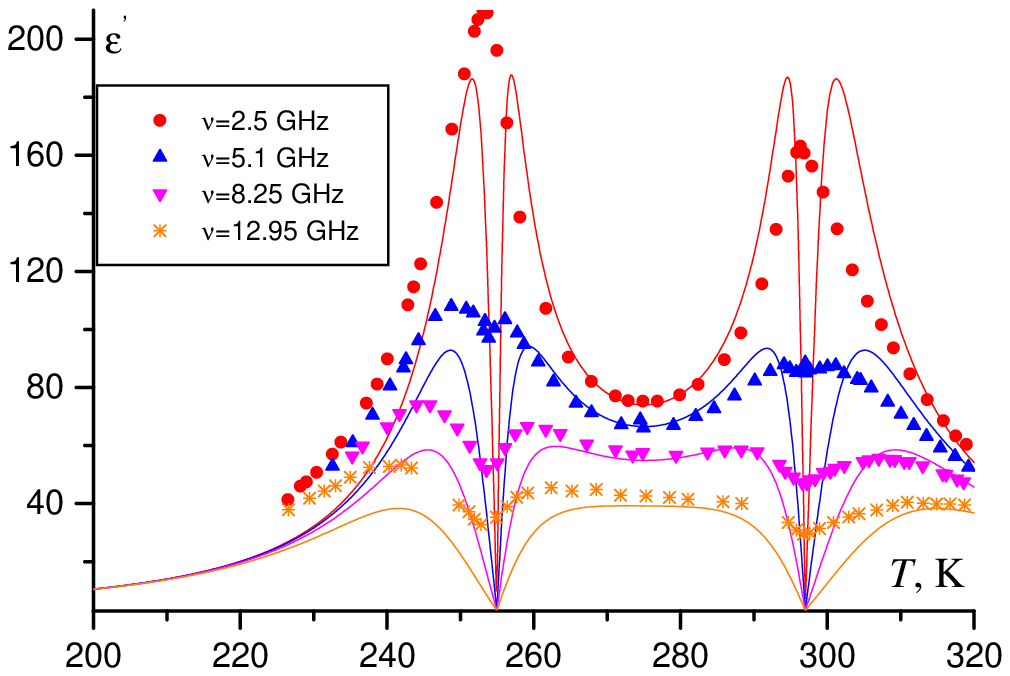}}
\end{center}
\caption{Temperature dependence of $\varepsilon'$ 
for Rochelle salt; lines are theoretical results for 
$\alpha=0.9\times10^{-13}$,
$\mu=1.8487\times10^{-18}$; symbols are experimental points
\protect\cite{sandy}.}
\end{figure}
\epsfysize=6.3cm
\begin{figure}
\begin{center}
\leavevmode
{\epsffile{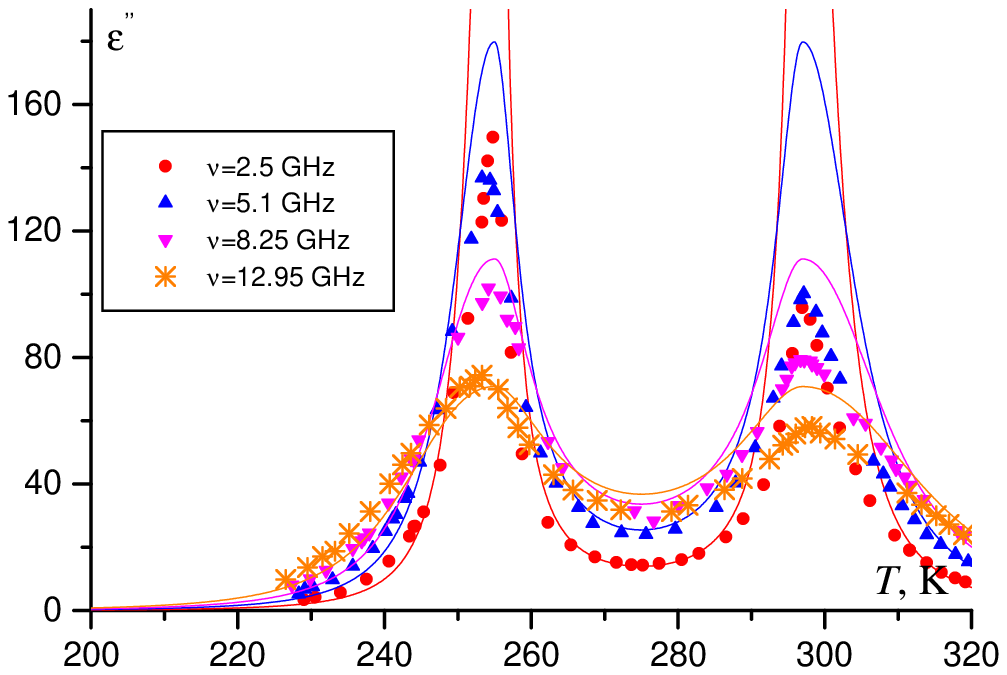}}
\end{center}
\caption{Temperature dependence of $\varepsilon''$ 
for Rochelle salt.}
\end{figure}
\epsfysize=13.5cm
\begin{figure}
\begin{center}
\leavevmode
{\epsffile{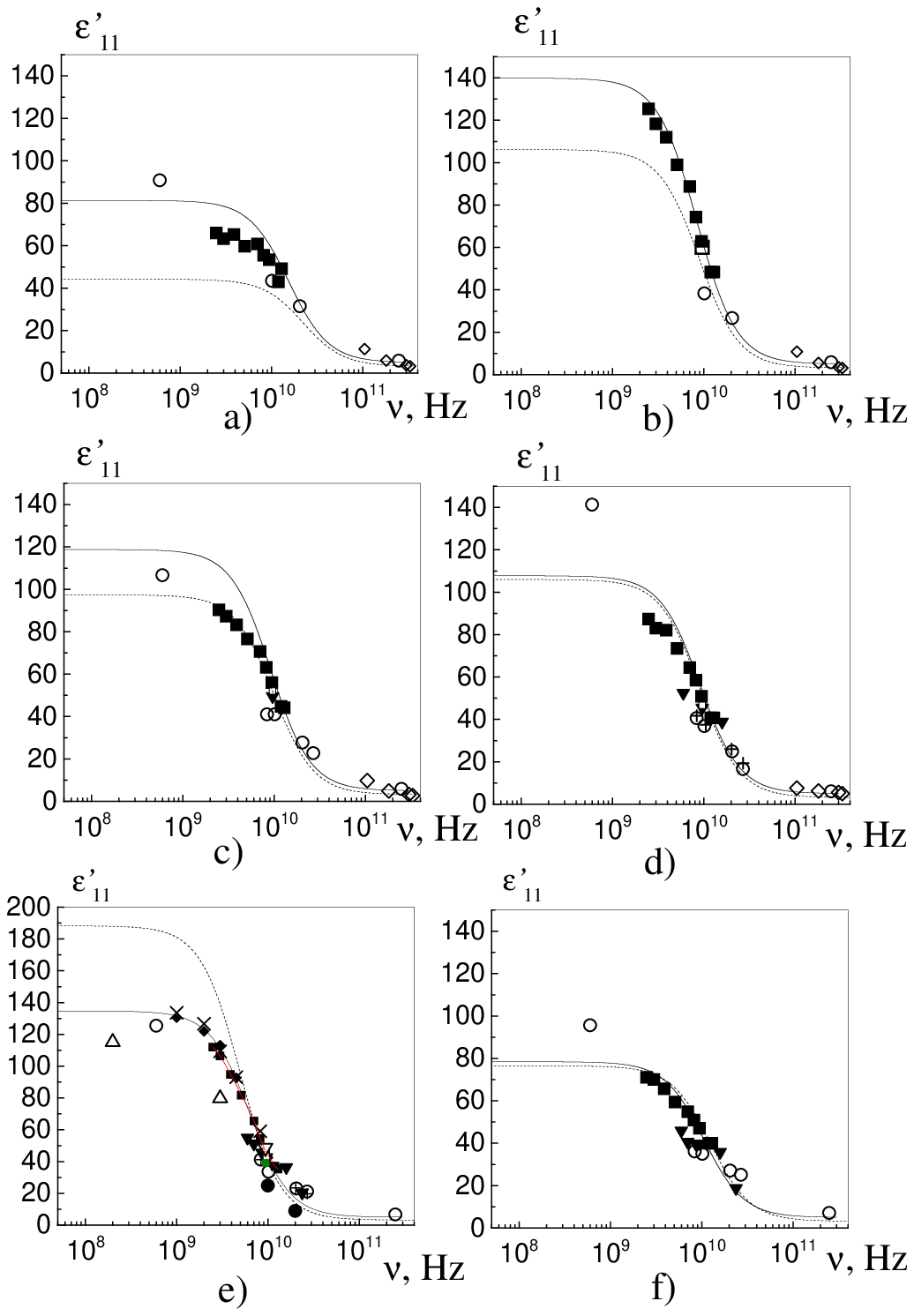}}
\end{center}
\caption{The frequency dependence of the real part of dynamic 
dielectric permittivity
 $\varepsilon'(\nu)$ at different temperatures $T$ (K):
 a) -- 235,
 b) -- 245,
 c) -- 265,
 d) -- 285,
 e) -- 305,
 f) -- 315.
Solid lines are the results from \protect\cite{alla}, dashed line is the 
result of present paper,
experimental points are taken from
 $\blacksquare$ -- \protect\cite{sandy},
 \protect\Large$\circ$\protect\small -- \protect\cite{61,62},
 $\blacktriangledown$ -- \protect\cite{66},
 $+$ -- \protect\cite{63},
 $\bullet$ -- \protect\cite{34},
 $\times$ -- \protect\cite{59},
 $\bigtriangleup$ -- \protect\cite{28},
 $\diamond$ -- \protect\cite{58}.
}
\end{figure}
\epsfysize=13.5cm
\begin{figure}
\begin{center}
\leavevmode
{\epsffile{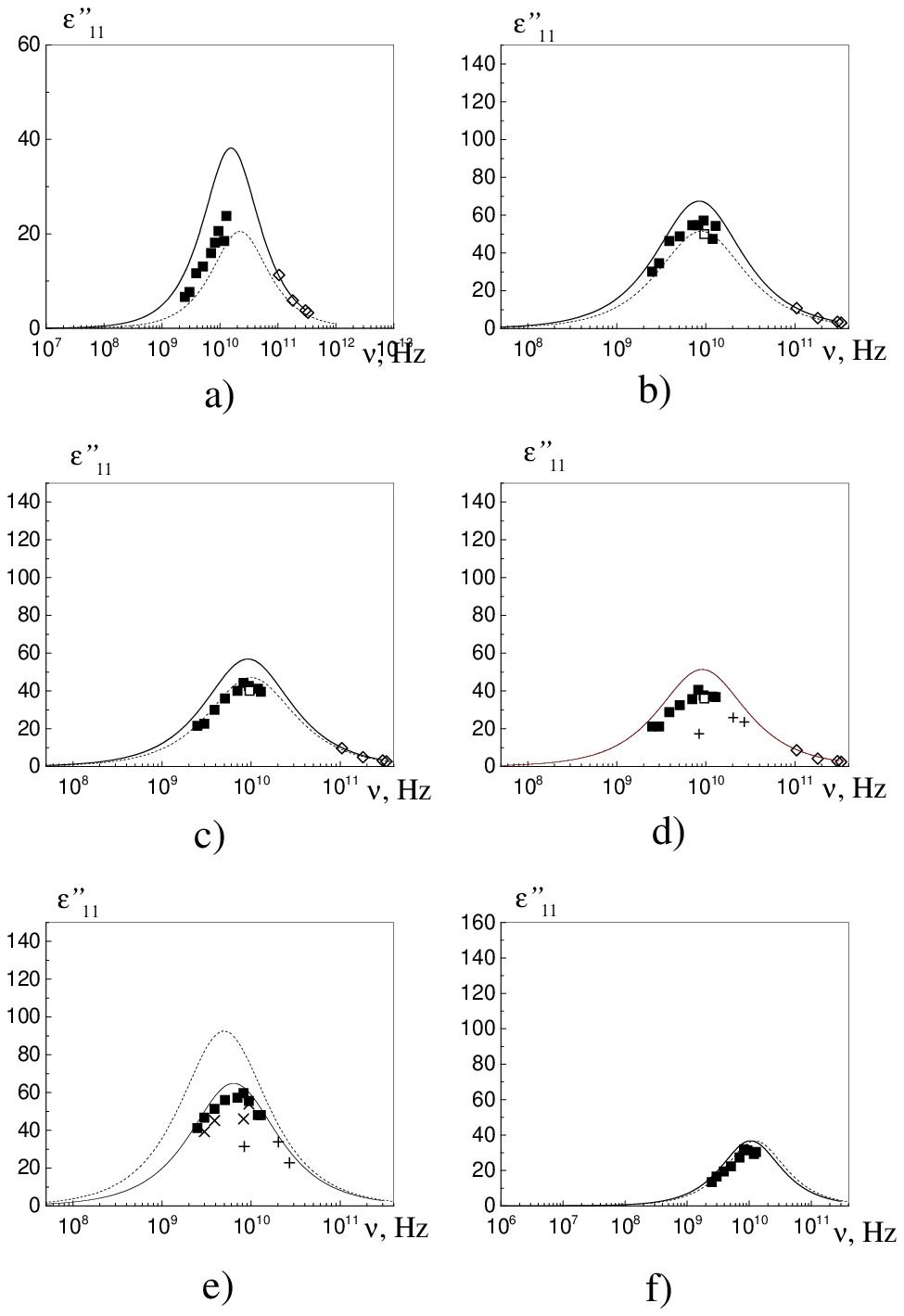}}
\end{center}
\caption{The frequency dependence of the imaginary part of dynamic 
dielectric permittivity
 $\varepsilon''(\nu)$ at different temperatures $T$ (K):
 a) -- 235,
 b) -- 245,
 c) -- 265,
 d) -- 285,
 e) -- 305,
 f) -- 315.
Solid lines are the results from \protect\cite{alla}, dashed line is the 
result of present paper,
experimental points are taken from
 $\blacksquare$ -- \protect\cite{sandy},
 \protect\Large$\circ$\protect\small -- \protect\cite{61,62},
 $\blacktriangledown$ -- \protect\cite{66},
 $+$ -- \protect\cite{63},
 $\diamond$ -- \protect\cite{60},
 $\times$ -- \protect\cite{59},
 $\square$ -- \protect\cite{64}.
 }
\end{figure}
\epsfysize=6.3cm
\begin{figure}
\begin{center}
\leavevmode
{\epsffile{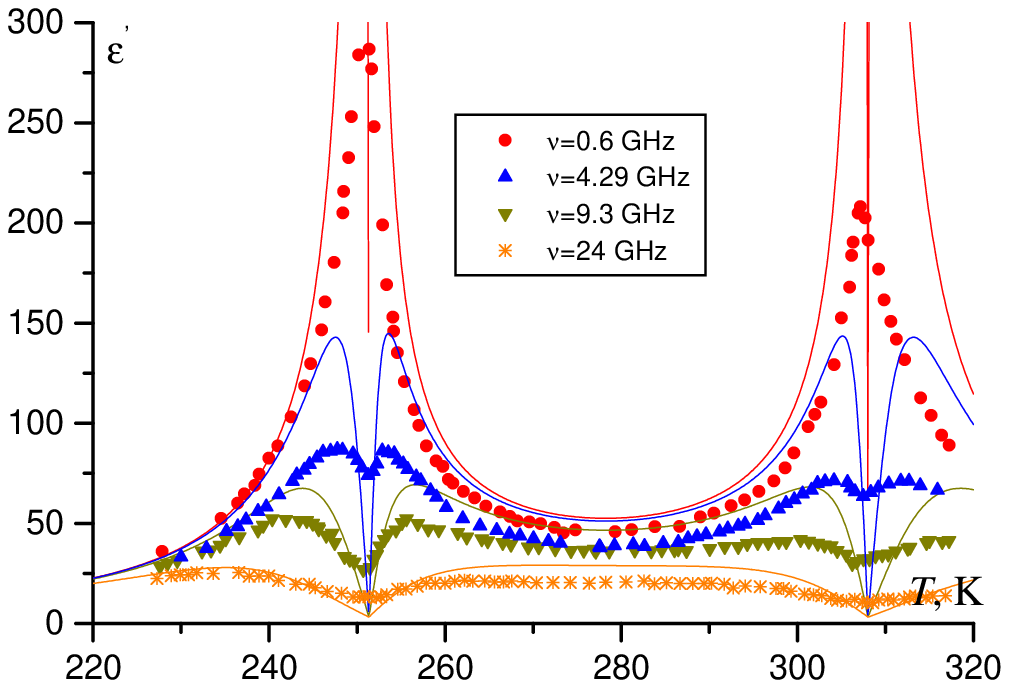}}
\end{center}
\caption{Temperature dependence of $\varepsilon'$ 
for deuterated Rochelle salt; lines are theoretical results for 
$\alpha=0.8\times10^{-13}$,
$\mu=2.00457\times10^{-18}$; symbols 
are experimental points \protect\cite{horioka}.}
\end{figure}
\epsfysize=6.3cm
\begin{figure}
\begin{center}
\leavevmode
{\epsffile{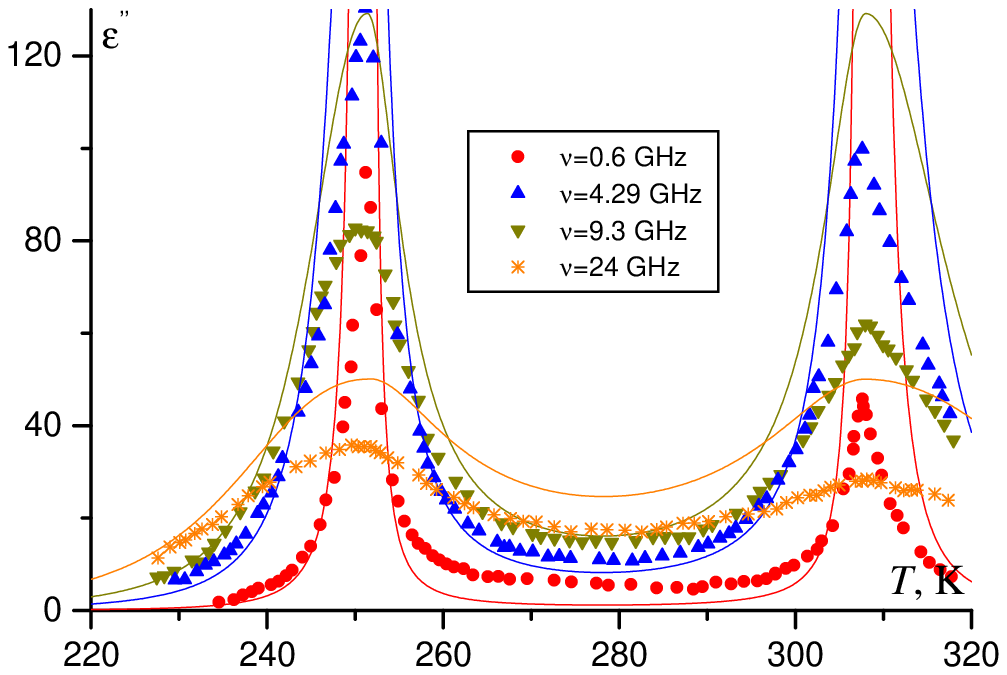}}
\end{center}
\caption{Temperature dependence of $\varepsilon''$ 
for deuterated Rochelle salt.}
\end{figure}
\epsfysize=6.0cm
\begin{figure}
\begin{center}
\leavevmode
{\epsffile{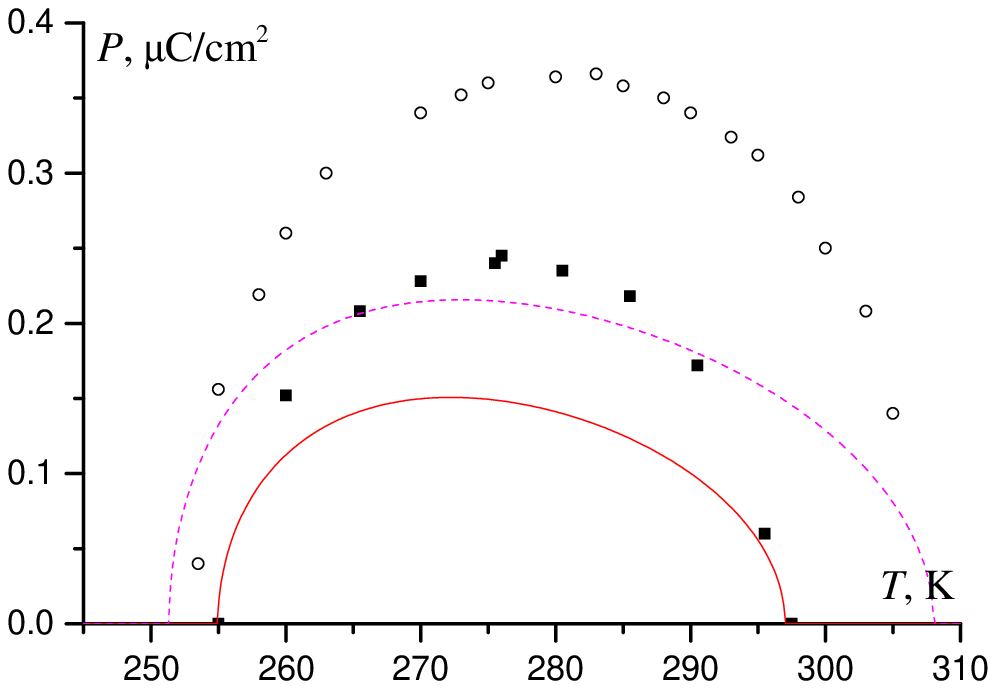}}
\end{center}
\caption{Temperature dependence of polarization of regular (solid line)
and deuterated (dashed line) Rochelle salt for sets of parameters 4, 5 
of table 4.1; $\mu_{Rs}=1.8487\times10^{-18}$, 
$\mu_{dRs}=2.0046\times10^{-18}$.}
\end{figure}
\epsfysize=6.0cm
\begin{figure}
\begin{center}
\leavevmode
{\epsffile{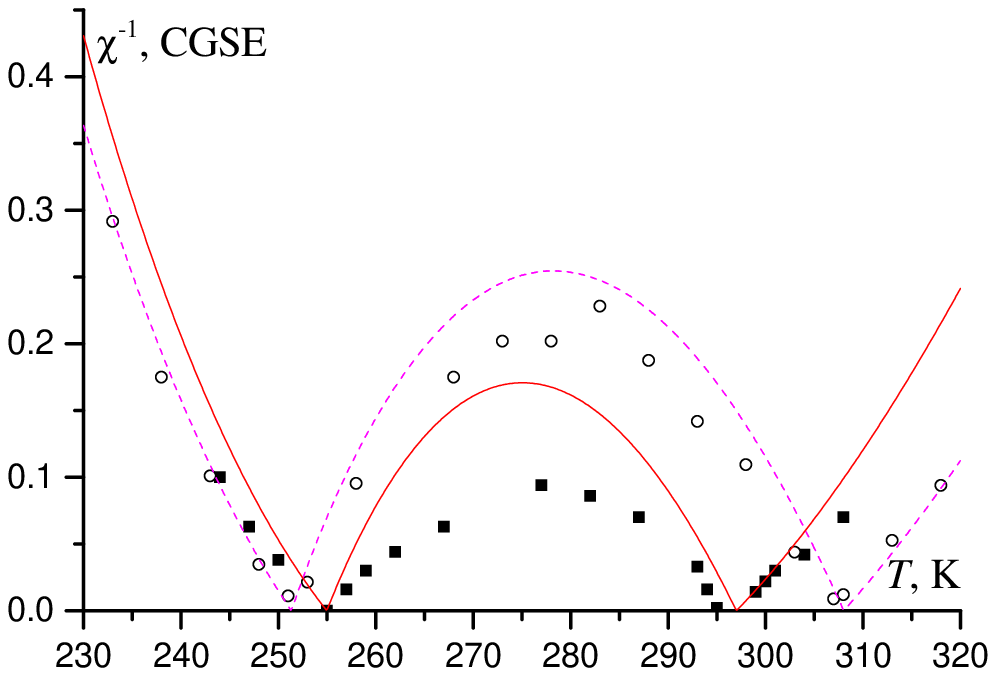}}
\end{center}
\caption{Temperature dependence of inverse static dielectric susceptibility 
of regular (solid line) and deuterated (dashed line) Rochelle salt for 
the same sets of parameters as in previous figure.}
\end{figure}
\epsfysize=6.3cm
\begin{figure}
\begin{center}
\leavevmode
{\epsffile{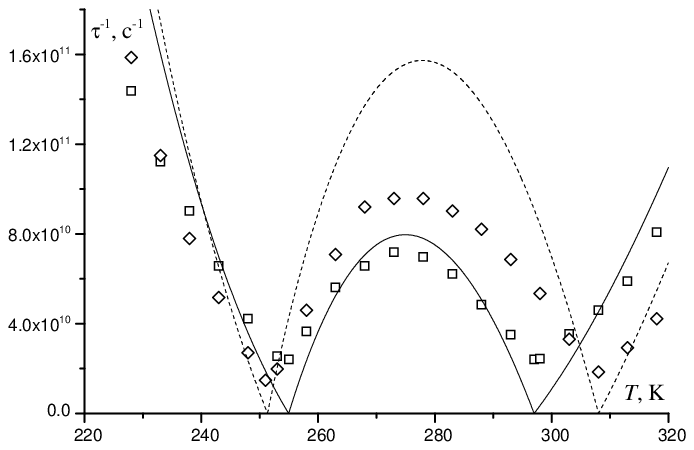}}
\end{center}
\caption{Temperature dependence of inverse relaxation times of regular 
(solid line) and 
deuterated (dashed line) Rochelle salt. Symbols are experimental points of
\protect\cite{horioka}).}
\end{figure}

\section{Thermodynamics and relaxational dynamics
of RbHSO$_4$.}
\setcounter{equation}{0}
\setcounter{table}{0}
\setcounter{figure}{0}
Ferroelectric properties of  RbHSO$_4$ were discovered in 1960 \cite{pep}.
This crystal undergoes a single second order
phase transition ($T_c=265$ K).
A microscopic theory of RbHSO$_4$ for the first time was proposed in
\cite{blat,alex}.
The studied model was based on the assumption that the phase transition in 
RbHSO$_4$ is related to ordering of sulphate groups,
which in a high-temperature phase move
in asymmetric double-well potentials. Hence, RbHSO$_4$ can be described within
the Mitsui model \cite{mitsui} (see (2.1)).
In
\cite{alex,ccc,blat,grigas,levit,levit2}
the thermodynamic and dielectric properties of RbHSO$_4$
RbDSO$_4$ and disordered mixtures RbH$_{1-x}$D$_x$SO$_4$ were studied. Unfortunately,
the found in these papers values of the microparameters do not provide a 
proper description of the experimentally observed jump of the specific
heat at the transition point ($\Delta c=9.02$ J/Mol K).

Since RbHSO$_4$ and related disordered mixtures
undergo a single second order phase transition, the possible values of the
model parameters for these compounds should be taken from the regions I, II 
of the phase diagram. Also, since the specific heat of the studied model
depends only on the dimensionless parameters  $a$ and $\gamma$, the values
of these parameters were  chosen such as to fit the theoretical jump
of specific heat to the experimental value. Then the parameters 
$K$, $J$, $\Delta$ were determined from $a$ and $\gamma$ using the condition
that the calculated transition temperature should coincide with
the experimental one.
Finally, for RbHSO$_4$ we obtained the following set of the parameters 
$K=880.64$ K, $J=780.9$ K, $\Delta=398.8$ K,
$\mu=0.425276\times 10^{-18}$.
Here we took into account the fact that for 
RbHSO$_4$ $v= 2.105\times 10^{-22}$ cm$^3$.

Thermodynamic and dielectric characteristics of the disordered 
compounds RbH$_{1-x}$D$_x$SO$_4$ were calculated in the mean 
field approximation under assumption that the corresponding 
model parameters are given by
\bea
K_x=K_H (1-x)+K_D x,
\nonumber\\
J_x=J_H (1-x)+J_D x,
\nonumber\\
\Delta_x=\Delta_H (1-x)+\Delta_D x.
\eea
Here $K_H$, $J_H$, $\Delta_H$ and  $K_D$, $J_D$, $\Delta_D$ 
are the model parameters, corresponding to RbHSO$_4$ and RbDSO$_4$, 
respectively. We took into account the fact that for RbDSO$_4$ $T_c=252$~K. 
Analogously to the case of RbHSO$_4$, we obtained  the values of the 
model parameters for RbH$_{0.3}$D$_{0.7}$SO$_4$, and then using (6.1)
found the parameters for RbDSO$_4$. Table 6.1 contains values of the model
parameters for RbHSO$_4$ and RbDSO$_4$, obtained in the present and 
previous \cite{alex,blat,grigas,levit} works. 
\begin{table}
\caption{The sets of the model parameters for RbHSO$_4$ 
and RbDSO$_4$, obtained within the mean field approximation (the sets
5 and 6 are found in the present work).}
\begin{tabular}{|c|l|l|l|l|l|l|}
\hline
No &compound                  &$K$, K &$J$, K&$\Delta$, K&$a$   &$\gamma$\\
\hline
1. &RbHSO$_4$ \cite{alex}    &616    &784   &245        &-0.12 &0.35    \\
2. &RbHSO$_4$ \cite{blat}    &1040   &228   &468        &~0.64 &0.738   \\
3. &RbHSO$_4$ \cite{grigas}  &616    &777.2 &244        &-0.116&0.35    \\
4. &RbHSO$_4$ \cite{levit}   &616    &784   &244        &-0.12 &0.348   \\
5. &RbHSO$_4$                &880.64 &780.9 &398.8      &~0.6  &0.48    \\
6. &RbDSO$_4$                &842.4  &747.1 &381.5      &~0.6  &0.48    \\
\hline
\end{tabular}
\end{table}
Using equation (6.1) and the presented in Table 6.1 values of the model 
parameters for RbHSO$_4$ and RbDSO$_4$, we calculated the physical 
characteristics of the mixed compounds
RbH$_{1-x}$D$_x$SO$_4$. The obtained results for polarization, inverse 
dielectric susceptibility, and contribution of the ordering structure 
elements to the specific heat of
RbH$_{1-x}$D$_x$SO$_4$ along with the available experimental data are shown 
in figs. 6.1-6.3. The fig.6.4 illustrates the calculated 
specific heat and the experimental points for RbHSO$_4$. In figs.6.5-6.9 
we plotted the calculated
$\varepsilon'(\nu,T)$, $\varepsilon''(\nu,T)$ for
RbHSO$_4$ and the inverse relaxation time
RbH$_{1-x}$D$_x$SO$_4$.
A good quantitative fit of the obtained theoretical results 
to experimental data for the studied characteristics of RbHSO$_4$ is 
obtained.
If figs.6.10-6.13 we show that the proposed theory also well 
describes the experimental data of
\cite{grigas} for temperature and frequency dependences of
$\varepsilon'(\nu,T)$, $\varepsilon''(\nu,T)$ for
RbH$_{0.3}$D$_{0.7}$SO$_4$.
Hence, a good agreement between the theoretical results and 
experimental data is obtained for thermodynamic and dynamic 
characteristics of the RbHSO$_4$ 
RbH$_{0.3}$D$_{0.7}$SO$_4$ ferroelectrics. We can maintain that the 
Mitsui model  (2.1) is totally adequate to these crystals.
\epsfysize=6.1cm
\begin{figure}
\begin{center}
\leavevmode
{\epsffile{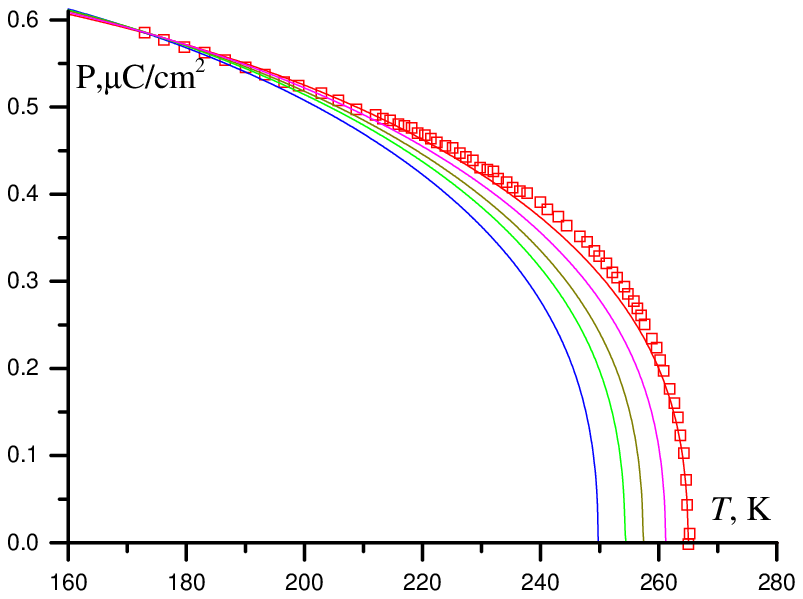}}
\end{center}
\caption{Temperature dependence of polarization of
RbH$_x$D$_{1-x}$SO$_4$ for $x=$1, 0.7, 0.5, 0.25, 0; $\square$ are the 
experimental points taken from \protect\cite{kajikawa}.}
\end{figure}
\epsfysize=6.1cm
\begin{figure}
\begin{center}
\leavevmode
{\epsffile{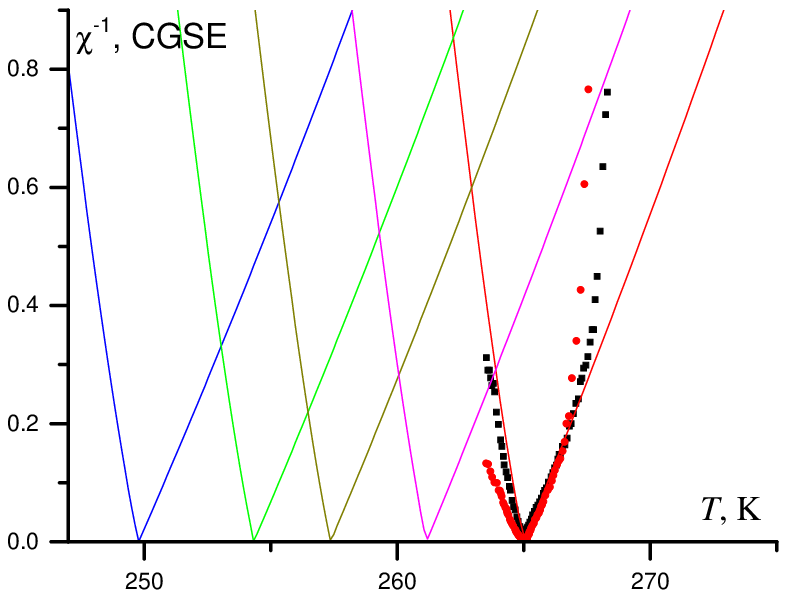}}
\end{center}
\caption{Temperature dependence of the inverse static 
susceptibility of
RbH$_x$D$_{1-x}$SO$_4$ for $x=$1, 0.7, 0.5, 0.25, 0; Symbols 
are the 
experimental points taken from  \protect\cite{kajikawa}.}
\end{figure}
\epsfysize=6.3cm
\begin{figure}
\begin{center}
\leavevmode
{\epsffile{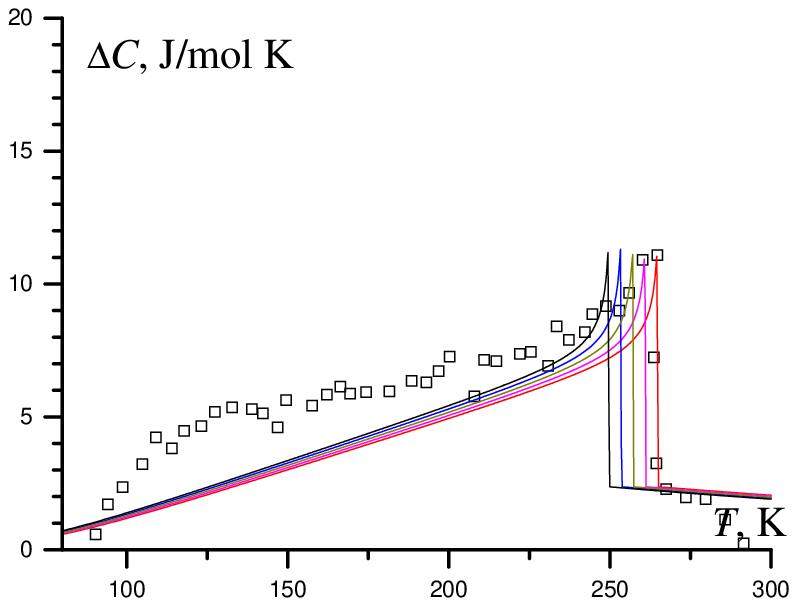}}
\end{center}
\caption{Contribution of the ordering elements to the specific heat  as 
a function of temperature
for RbH$_x$D$_{1-x}$SO$_4$ at $x=$1, 0.7, 0.5, 0.25, 0.}
\end{figure}
\epsfysize=6.3cm
\begin{figure}
\begin{center}
\leavevmode
{\epsffile{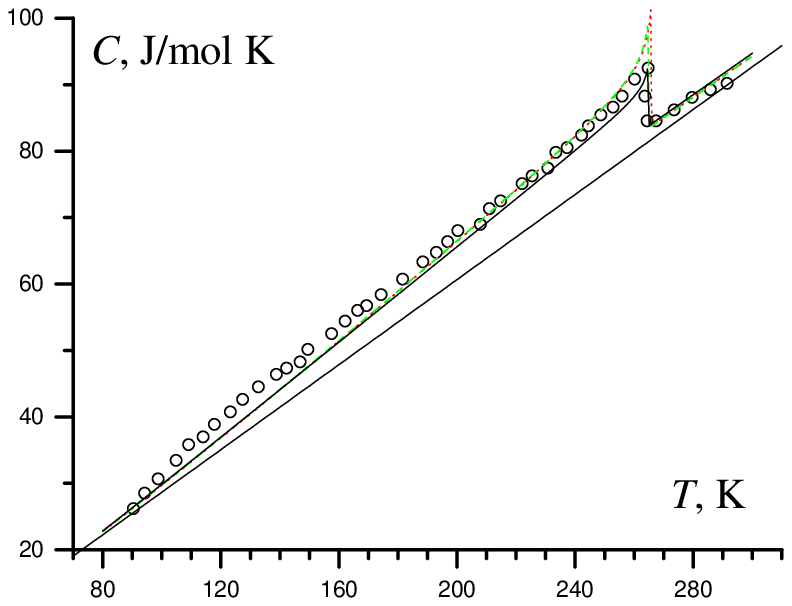}}
\end{center}
\caption{Specific heat of RbHSO$_4$ as a function of temperature 
(the solid line represents theoretical results obtained within the 
mean field approximation, the symbols are experimental points 
\protect\cite{alex}).}
\end{figure}
\epsfysize=6.3cm
\begin{figure}
\begin{center}
\leavevmode
{\epsffile{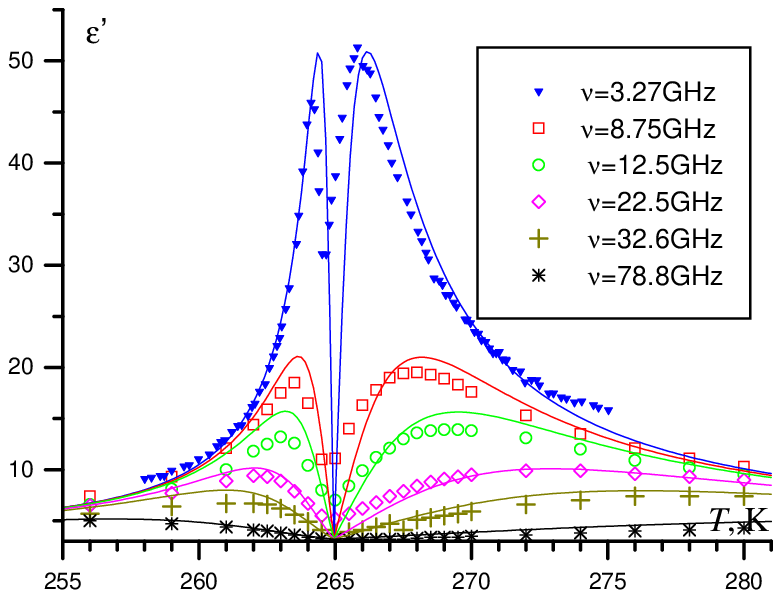}}
\end{center}
\caption{Temperature dependence of $\varepsilon'$ 
of RbHSO$_4$
($\alpha=0.48\times 10^{-13}$):  the solid line represents theoretical 
results; the symbols are experimental points 
\protect\cite{grigas}.}
\end{figure}
\epsfysize=6.3cm
\begin{figure}
\begin{center}
\leavevmode
{\epsffile{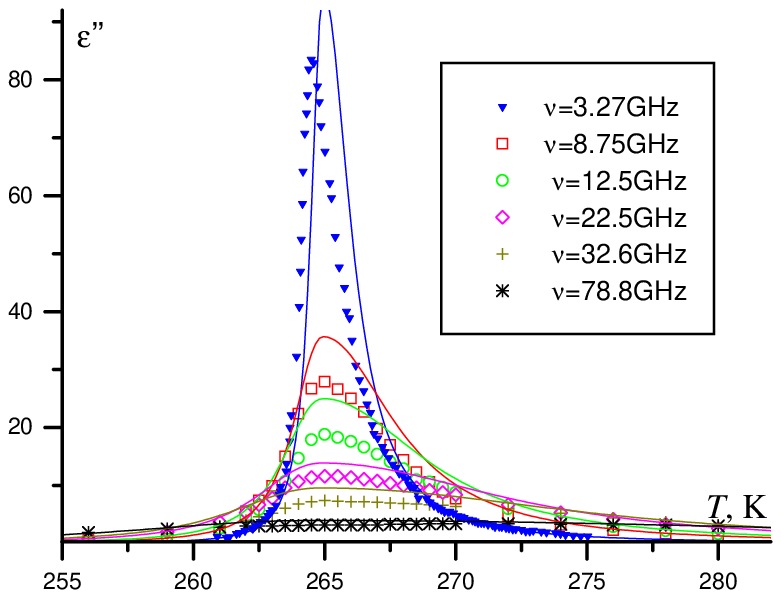}}
\end{center}
\caption{Temperature dependence of $\varepsilon''$ 
of RbHSO$_4$
($\alpha=0.48\times 10^{-13}$):  the solid line represents theoretical 
results; the symbols are experimental points \protect\cite{grigas}.}
\end{figure}
\epsfysize=6.3cm
\begin{figure}
\begin{center}
\leavevmode
{\epsffile{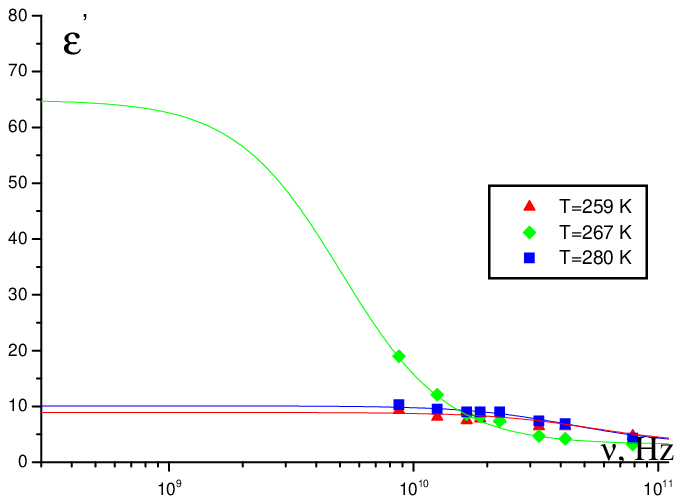}}
\end{center}
\caption{$\varepsilon'$ of RbHSO$_4$ as a function of frequency;
the solid lines represent theoretical 
results; the symbols are experimental points \protect\cite{grigas}.}
\end{figure}
\epsfysize=6.3cm
\begin{figure}
\begin{center}
\leavevmode
{\epsffile{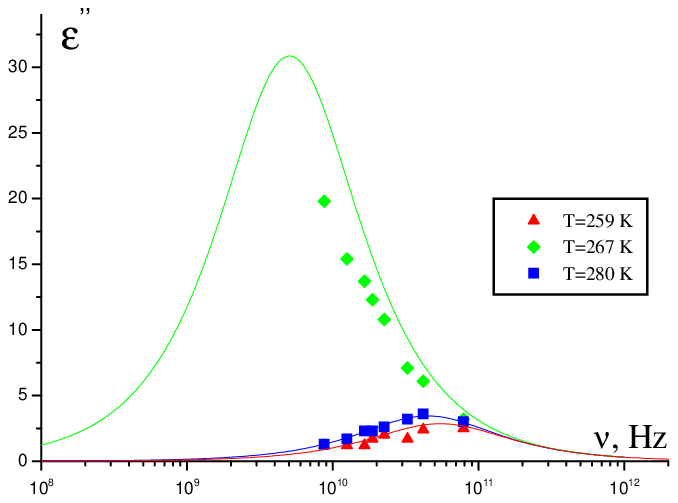}}
\end{center}
\caption{$\varepsilon''$ of RbHSO$_4$ as a function of frequency;
the solid lines represent theoretical 
results; the symbols are experimental points \protect\cite{grigas}.}
\end{figure}
\epsfysize=6.3cm
\begin{figure}
\begin{center}
\leavevmode
{\epsffile{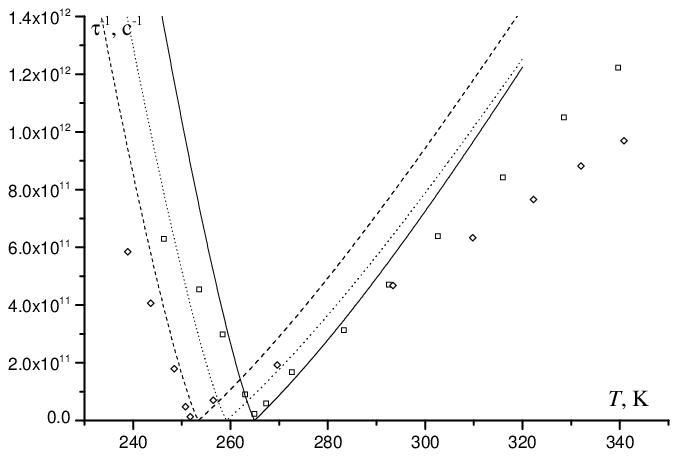}}
\end{center}
\caption{Temperature dependence of the inverse relaxation times 
of RbHSO$_4$
(solid line) and RbDSO$_4$ (dashed line),
$\square$ - \protect\cite{grigas}, $\diamond$ - \protect\cite{ftt}.}
\end{figure}
\epsfysize=6.3cm
\begin{figure}
\begin{center}
\leavevmode
{\epsffile{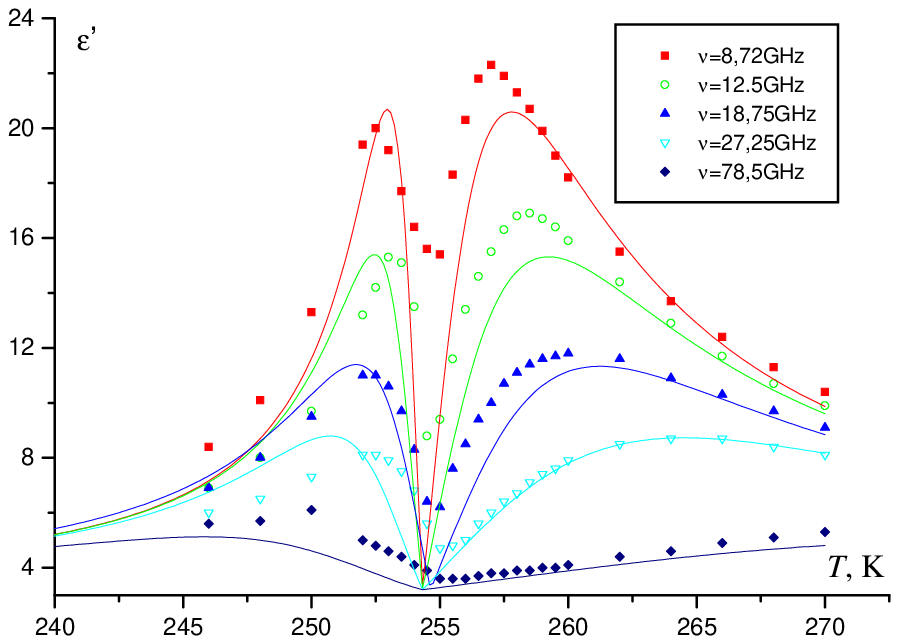}}
\end{center}
\caption{Temperature dependence of $\varepsilon'$ for
RbH$_{0.3}$D$_{0.7}$SO$_4$
($\alpha=0.54\times 10^{-13}$): the solid lines represent theoretical 
results; the symbols are experimental points \protect\cite{grigas}.}
\end{figure}
\epsfysize=6.3cm
\begin{figure}
\begin{center}
\leavevmode
{\epsffile{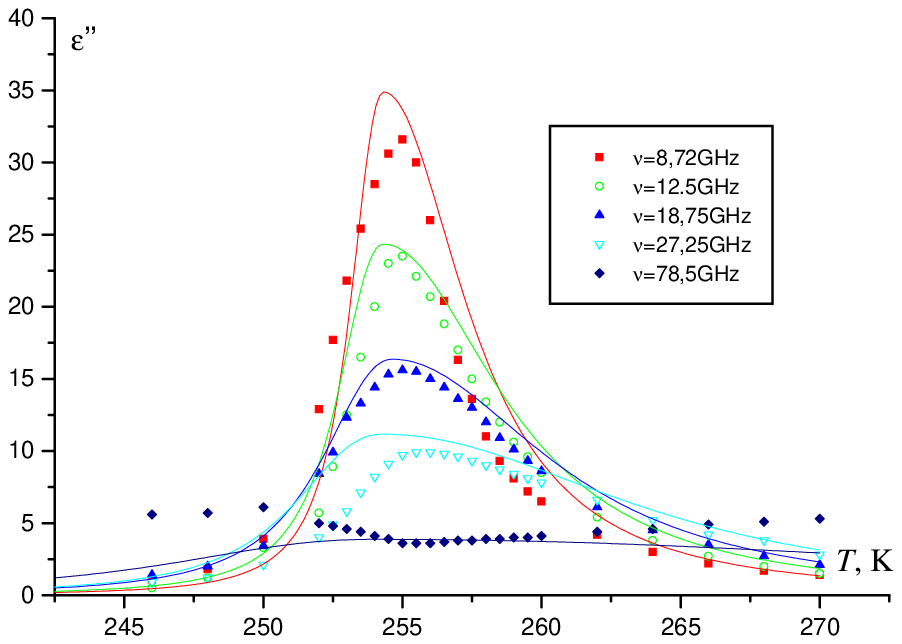}}
\end{center}
\caption{Temperature dependence of $\varepsilon''$  for
RbH$_{0.3}$D$_{0.7}$SO$_4$
($\alpha=0.54\times 10^{-13}$): the solid lines represent theoretical 
results; the symbols are experimental points \protect\cite{grigas}.}
\end{figure}
\epsfysize=6.3cm
\begin{figure}
\begin{center}
\leavevmode
{\epsffile{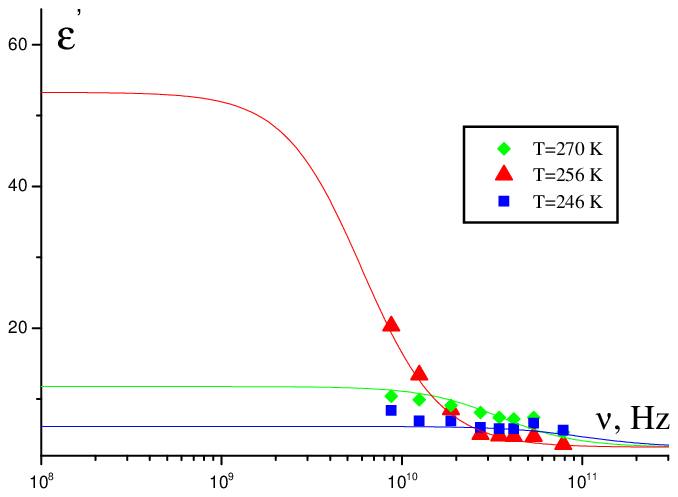}}
\end{center}
\caption{$\varepsilon'$ of RbH$_{0.3}$D$_{0.7}$SO$_4$ as a function of 
frequency; the solid lines represent theoretical 
results; the symbols are experimental points \protect\cite{grigas}.}
\end{figure}
\epsfysize=6.3cm
\begin{figure}
\begin{center}
\leavevmode
{\epsffile{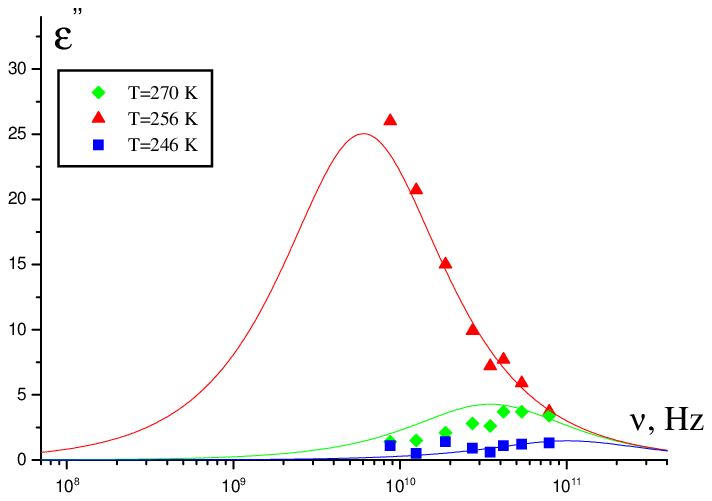}}
\end{center}
\caption{$\varepsilon''$ RbH$_{0.3}$D$_{0.7}$SO$_4$ as a function of 
frequency; the solid lines represent theoretical 
results; the symbols are experimental points \protect\cite{grigas}.}
\end{figure}

\section{Concluding remarks}

In this paper we calculated thermodynamic and dynamic characteristics of 
ferroelectric order-disorder type compounds described by the pseudospin 
model with asymmetric double-well potential. For the first time, a 
role of tunneling on the systems described by this model is studied 
thoroughly; its influence on the phase diagram is shown. Possible phase 
transitions  are analysed at different values of the model  parameters; at 
these values the physical characteristics of the model are calculated. 
Interesting results are obtained for compounds undergoing three phase 
transitions, like
\{(NH$_4$)$_3$H(SO$_4$)$_2$\}$_{1-x}$\{(ND$_4$)$_3$D(SO$_4$)$_2$\}$_x$
\cite{osaka}. Because of a limited number of experimental 
data available for these crystals, we cannot draw final conclusions 
about the model adequacy for these crystals. We also calculated the physical 
characteristics of the  $(dRs)_xRs_{1-x}$ and RbH$_{1-x}$D$_x$SO$_4$ type 
ferroelectrics.

Our study has shown that within the studied model it is most likely 
impossible to describe the phase transition and physical characteristics of
NH$_4$HSO$_4$ crystal. Interesting results are obtained for the  RbHSO$_4$ 
and $Rs$ type ferroelectrics. The first crystal is not piezoelectric in 
the paraelectric phase, whereas  the piezoelectric effect 
determines the physical characteristics of the second one \cite{alla}. 
The considered 
here model does not take into account the piezoelectric interaction.

It should be also noted that in the mentioned ferroelectrics, an interaction 
of the ordering structure elements with phonons is rather important. 
Corresponding studies for these crystals were performed in
\cite{levsor,phonon1,phonon2}.

Hence, despite a great attention to the  crystals described by the Mitsui 
model, especially to Rochelle salt and of RbHSO$_4$ type, it still 
unclear whether this model provides a quantitative description of the entire 
spectrum of physical characteristics for these crystals. Indeed, 
in practically all earlier paper, the microparameters were chosen by 
fitting to only a few characteristics of the crystals, whereas other 
quantities were not calculated with these parameters.  Naturally, 
the incorrectly chosen values of the $K$, $J$, $\Delta$ parameters 
led to the incorrect values of $\mu$ and $\alpha$ as well.
The aim of the  present paper was to calculate all physical 
characteristics of the Mitsui model, study their dependences on the 
microparameters, explore the tunneling effects on these characteristics. We 
also intended to develop a fitting procedure  for regular and deuterated 
Rochelle salt and
RbHSO$_4$ type ferroelectrics, which would provide 
a good fit to the available 
experimental data for these crystals and help to establish the adequacy (or 
inadequacy) of the considered model to the specific crystals.

In \cite{lisnyi} it has been shown that an important characteristics in 
the fitting procedure for ferroelectric materials is a specific 
heat. In the present paper at setting the microparameters $K$, $J$, $\Delta$ 
we used the data for the specific heat of the studied crystals. However, it 
should be mentioned that for specific heat $Rs$ and $dRs$ reliable data 
are most likely absent.

The performed calculations yielded the values of the microparameters for 
the RbHSO$_4$ crystals, with which we calculated the physical 
characteristics of the disordered ferroelectrics RbH$_{1-x}$D$_x$SO$_4$.
The proposed consistent fitting scheme provides a good 
quantitative description of experimental data for pure crystals; we 
also present the results for disordered crystals RbH$_{1-x}$D$_x$SO$_4$
which can be experimentally verified.

For $Rs$ and $dRs$ crystals it is impossible to find such values of 
the model parameters, which would provide a simultaneous fit to the 
experimental data for $P_s(T)$ and $\varepsilon(0,T)$. An agreement 
could be somewhat improved by taking into account the piezoelectric 
interaction. This is also indicated by calculations of the dynamic 
characteristics of the crystals for which a satisfactory agreement 
with experimental data is obtained. It is clear that with increasing 
frequency, the effects of piezoelectric interaction in $Rs$ type crystals 
become unimportant. The crystals become clamped, and we obtain a good 
agreement with experimental data for $\varepsilon'(\nu,T)$ and 
$\varepsilon''(\nu,T)$.

Another important point is, if we choose the values of the microparameters 
$K$, $J$, $\Delta$ at which (see the phase diagram) the third 
low-temperature phase transition in  $Rs$ emerges, an agreement with the 
experimental data simultaneously for $P_s(T)$ and $\varepsilon(0,T)$ could 
be much improved. However,  the question about the 
low-temperature transition in $Rs$ is still open, even though such a 
transition was reported in some papers.

\end{document}